\documentclass[a4paper,8pt]{revtex4}
\usepackage{graphicx}  
\usepackage{amsmath}   
\usepackage{amssymb}   
\usepackage{bm} 
\usepackage{dcolumn}
\usepackage{color}
\usepackage{mathrsfs}
\usepackage{amsfonts}
\usepackage{varioref}
\usepackage{mathrsfs}
\usepackage{graphicx}
\usepackage{latexsym}
\usepackage{amsmath}
\usepackage{amssymb}
\usepackage{textcomp}
\usepackage{amsbsy}
\usepackage{graphics}
\usepackage{epstopdf}
\usepackage{color}
\usepackage[caption=false]{subfig}

\RequirePackage[colorlinks,citecolor=blue,urlcolor=magenta,linkcolor=blue]{hyperref}
\input epsf

\begin{document}

\tolerance=5000

\title{Aspects of non-singular bounce in modified gravity theories}

\author{Indrani~Banerjee$^{1}$\,\thanks{banerjeein@nitrkl.ac.in},
Tanmoy~Paul$^{2}$\,\thanks{pul.tnmy9@gmail.com},
Soumitra~SenGupta$^{3}$\,\thanks{tpssg@iacs.res.in}} \affiliation{
$^{1)}$ Department of Physics and Astronomy, National Institute of Technology, Rourkela-769008, India\\
$^{2)}$ Department of Physics, Chandernagore College, Hooghly - 712 136, India\\
$^{3)}$ School of Physical Sciences, Indian Association for the Cultivation of Science, Kolkata-700032, India}


\tolerance=5000

\begin{abstract}
Scenario of a bouncing universe is one of the most active area of research to arrive at singularity free cosmological models. 
Different proposals have been suggested to avoid the so called 'big bang' singularity through the quantum aspect of gravity 
which is yet to have a proper understanding. In this work, on the contrary, 
we consider three different approaches, each of which goes beyond General Relativity but remain within the domain of classical cosmological scenario, 
to address this problem. The hallmark of all these approaches is that the origin of the bouncing mechanism is somewhat natural 
within the geometrical framework of the model without any need of incorporating external source by hand. 
In the context of these scenarios ,we also discuss various constraints that these viable cosmological models need to satisfy.

\end{abstract}

\maketitle

\section{Introduction}
One of the major challenges in cosmology is to resolve  the problem of an initial singularity also known as the Big-Bang singularity. 
The proposal of a  non-singular bounce 
leading to a possible singularity free expansion of the universe is therefore a subject of great interest.
Various alternative models of gravitational theories can be possible testing grounds to look for such a feature in the evolution of the Universe.
Some of the early universe scenarios, proposed so far, that can generate an almost scale invariant power spectrum and hence , 
confront the observational constraints 
are the \emph{inflationary scenario} \cite{guth,Linde:2005ht,Langlois:2004de,Riotto:2002yw,Baumann:2009ds}, 
the \emph{bouncing universe} \cite{Brandenberger:2012zb,Brandenberger:2016vhg,Battefeld:2014uga,Novello:2008ra,Cai:2014bea,Cai:2016thi,
deHaro:2015wda,Lehners:2008vx,Cai:2013kja,Odintsov:2015zza,Odintsov:2020fxb,Martin:2001ue,Buchbinder:2007ad,Peter:2002cn,Gasperini:2003pb,Creminelli:2004jg,
Cai:2014xxa,Odintsov:2020zct,Cai:2010zma,Avelino:2012ue,Barrow:2004ad,Haro:2015zda,Elizalde:2014uba,
Finelli:2001sr,Raveendran:2017vfx,Raveendran:2018yyh,Qiu:2010ch,Elizalde:2020zcb}, 
the \emph{emergent universe scenario} \cite{Ellis:2003qz,Paul:2020bje,Bag:2014tta} 
and the \emph{string gas cosmology} \cite{Brandenberger:2008nx,Brandenberger:2011et,Battefeld:2005av}. 

In this review article, we are interested to explore the bounce cosmology from various perspectives in modified theories of gravity. 
Although it is believed 
that the Big-Bang singularity may be avoided through a suitable quantum generalization of gravity, 
the absence of a consistent quantum theory of gravity makes the bouncing description a strong area of interest. However 
the bounce cosmology are hinged with some serious issues, like:
\begin{itemize}
\item The bounce model(s) generally predict a large value of tensor to scalar ratio (compared 
to the Planck constraint \cite{Akrami:2018odb}) in the 
perturbation evolution, i.e., the scalar and tensor perturbations get comparable amplitudes to each other \cite{Brandenberger:2016vhg}. 

\item The spacetime anisotropic energy density seems to grow faster than that of the bouncing agent during the contracting phase, 
which in turn makes the background evolution unstable (also known as BKL instability) \cite{new1}.
 
\item The Hubble radius in bounce scenario generally increases with time after the bounce, i.e., the bounce model(s) 
are unable to explain the late time acceleration or 
equivalently the dark energy era of universe, which, 
in fact, is not consistent with the recent supernovae observations indicating a current accelerating stage of universe 
\cite{Perlmutter:1998np}.
\end{itemize}

Here we mainly concentrate on the above problems and their possible resolutions in three different models of gravity beyond
Einstein’s General Relativity \cite{Banerjee:2020uil,Paul:2022mup,Odintsov:2021yva}. Here we would like to mention that 
even though some bounce models do predict large values of the tensor-to-scalar ratio ($r$), 
certain bouncing scenarios have been discussed in the literature 
(for example, \cite{Raveendran:2017vfx,Raveendran:2018why}) which lead to $r$ values compatible with Planck constraints.

In general, the modifications of Einstein’s gravity can be broadly classified as -- (1) 
by introducing some
extra spatial dimension over our usual four dimensional spacetime 
\cite{Csaki:2004ay,Brax:2003fv}, 
by inclusion of higher curvature term in addition to the Einstein–Hilbert term in the gravitational
action \cite{Nojiri:2010wj,Capozziello:2011et}, and (3) by introducing additional matter fields. 

In Sec.~[\ref{sec-b1}] we consider a non-flat warped braneworld model to examine a viable bounce scenario that predicts a low tensor-to-scalar ratio 
over the large scales and becomes consistent with the Planck data. 
It is well known that various extra dimensional models can provide a plausible resolution to the 
gauge-hierarchy/ finetuning problem in particle physics arising due to large radiative  corrections of the 
Higgs mass \cite{Csaki:2004ay,ArkaniHamed:1998nn,Randall:1999ee}. 
In particular, the warped geometry models due to 
Randall-Sundrum (RS) \cite{Randall:1999ee} earned a lot of attention since it resolves the gauge hierarchy problem without invoking 
any intermediate scale (between Planck and TeV scale) in the theory.
The assumption of flat branes in RS scenario can be relaxed in a generalized warped braneworld model \cite{Das:2007qn}, 
which allows the branes to be non-flat giving rise to de Sitter (dS) or anti-de Sitter (AdS) branes. The cosmological, astrophysical and phenomenological 
implications of warped braneworld models (with flat or non-flat branes) have been discussed in 
\cite{Csaki:1999mp,Binetruy:1999ut,Davoudiasl:1999jd,Paul:2016itm}. 
In the non-flat warped braneworld scenario the non-vanishing vacuum energy on the brane 
enables the radion ( the field originating from the interbrane separation ) to generate its own potential along with 
a non-canonical kinetic term in the four 
dimensional effective action which in turn can stabilize the modulus to the suitable value \cite{Banerjee:2017jyk}, 
without invoking any additional scalar field in the theory.
This  non-canonical scalar kinetic term becomes negative for certain values of the modulus which endows the 
radion a phantom-like behavior where the null energy condition is violated. This, being a generic 
feature  in a bouncing universe,  motivates us to explore the prospect of the non-flat warped 
braneworld model in addressing  a viable bouncing cosmology. We investigate the cosmological evolution of the radion 
field in the FRW background and the subsequent evolution of the primordial fluctuations which allows us to 
understand the viability of the model in purview of the Planck 2018 constraints.\\

On a different side, it is well known that  the present universe bears  the signatures of rank two symmetric 
tensor field in the form of gravity, while  it carries no noticeable presence of rank two antisymmetric tensor field, generally known as Kalb-Ramond field 
\cite{Kalb:1974yc}. 
Such KR field also arise as closed string mode, and are of considerable interest, 
in the 
context of String theory.  In the arena of higher dimensional braneworld scenario or in higher curvature gravity theory, 
it has been shown that the KR 
coupling (with other normal matter fields) is highly suppressed over the usual gravity-matter coupling, 
which explains why the KR field has not been detected so far \cite{Mukhopadhyaya:2002jn,Das:2018jey}. 
However the KR field is expected to show its considerable effect during the early stage of the universe 
\cite{Elizalde:2018rmz,Aashish:2018lhv}. 
In this context it has been shown that
the presence of KR field slows down the acceleration of the universe and also modifies the inflationary observable parameters 
\cite{Elizalde:2018rmz}. 
For example the presence of the KR field in a higher curvature $F(R) = R+R^3$ inflationary model 
is a viable model in respect to the Planck data, unlike  the case without the KR field \cite{Elizalde:2018rmz}. 
Motivated by this, in Sec.~[\ref{sec-b2}], 
we explore the possible roles of KR field in driving a non-singular bounce in particular an $ekpyrotic~bounce$ 
driven by the second rank antisymmetric Kalb-Ramond field in F(R) gravity theory. \\

Finally in Sec.~[\ref{sec-b3}], we aim to study a smooth unified scenario from a non-singular bounce to the dark energy era. 
For this , here we consider the Chern-Simons (CS) corrected F(R) gravity theory, where the presence of the CS coupling induces a 
parity violating term in the gravitational action. Such  term arises naturally  in the low-energy effective action of several string inspired 
models \cite{Green:1984sg,Antoniadis:1992sa} . and it may have important role in explaining 
the primordial power spectrum and may possibly provide an indirect testbed for string theory. 
Furthermore the parity violating Chern-Simons gravity distinguishes the evolution of the two polarization modes of primordial gravitational 
waves \cite{Hwang:2005hb}, which leads to the generation of 
chiral gravitational waves leaving non-trivial imprints in the Cosmic 
Microwave Background Radiation (CMBR). Such signatures, if detected in the future generation of experiments, may signal the presence of the string 
inspired Chern-Simons gravity in the early universe. 
The astrophysical implications of the gravitational Chern-Simons (GCS) term has been explored e.g. \cite{Wagle:2018tyk}.
Another important motivation to include the CS term in the context of F(R) gravity 
originates  from the fact that it can correctly reproduce the observed tensor-to-scalar ratio in 
respect to the Planck data, which is otherwise not possible in the context of $F(R)$ gravity model only \cite{Odintsov:2019mlf}. 
Actually the CS term does not affect the evolution of the spatially flat FRW background or the scalar perturbations, 
but plays crucial role in  tensor perturbations, which in turn reduces the amplitude of the tensor perturbation compared to that 
of the scalar perturbation. 
All these motivate us to explore the relevance of such a 
model in inducing a $bouncing$ universe and subsequently unifying it with the dark energy epoch. Based on these arguments, in the present paper, 
we try to provide a cosmological model which smoothly unifies certain cosmological era of the universe, particularly 
from a non-singular bounce to a matter dominated era and from the matter dominated to the dark energy epoch. 

The following notations are used in the paper: $t$ is the cosmic time, $\eta$ represents the conformal time defined by $d\eta = dt/a(t)$ with $a(t)$ being 
the scale factor of the universe. An overdot denotes the derivative with respect to cosmic time, while a prime symbolizes $\frac{d}{d\eta}$.

\section{\underline{Bounce from curved braneworld}}\label{sec-b1}

\subsection{The model}
\label{S2}

This section is organized as follows: in Sec.[\ref{S2}], we briefly describe the non-flat warped braneworld model and its four dimensional effective theory. 
Having set the stage, Sec.[\ref{S3}] is dedicated for studying the background cosmological evolution while the evolution of the perturbations and 
confrontation of the theoretical predictions with the latest Planck observations is discussed in Sec.[\ref{S4}]. A summary of 
our results are explained.\\
In the context of Randall-Sundrum (RS) 5 dimensional warped braneworld scenario with two 3-branes embedded within the full spacetime, the 
bulk metric is given by,
\begin{align}
{ds}^2=e^{-2A(r_c, \phi)}g_{\mu \nu}(x) {dx}^{\mu}{dx}^{\nu} + r_c ^2 {d\phi}^2~~. 
\label{Eq1}
\end{align}
The extra dimension coordinated by $\phi$ is $S^1/Z_2$ compactified, $r_c$ is the interbrane separation, $\{x^{\mu}\}$ are the brane coordinates 
and $A(r_c,\phi)$ is known as the warp factor. In the original RS scenario, the branes are considered to be flat, i.e $g_{\mu\nu}$ is replaced by the 
Minkowski metric $\eta_{\mu\nu}$, and the warp factor is obtained by solving the 5 dimensional Einstein equations as $A(r_c,\phi) = k_0r_c\phi$. Here 
$k_0=\sqrt{-\Lambda/24M^3}$ 
such that $\Lambda$ and $M$ denote the five dimensional cosmological constant and Planck mass respectively. In the case of flat branes, the bulk cosmological 
constant and the brane tensions are finely adjusted so that the brane curvature gets identically zero. In tis regard, 
one of our authors proposed a more general solution by relaxing the assumption of $flat$ branes and considered the branes 
to be either de-Sitter or anti de-Sitter characterized by a positive or negative brane cosmological constant respectively. In the case 
of non-flat braneworld scenario, induced on-brane metric is the solution of $G_{\mu\nu} = -\Omega g_{\mu\nu}$ where $G_{\mu\nu}$ is the Einstein 
tensor formed by $g_{\mu\nu}$ and $\Omega$ is the brane cosmological constant ($\Omega > 0$ for dS branes and $\Omega < 0$ for AdS branes). 
The warp factor is given by,
\begin{align}
e^{-A}=\omega \sinh\left(\ln\frac{c_2}{\omega}-k_0r_c |\phi| \right)~~~~~\mathrm{for~dS~brane}\nonumber\\
e^{-A}=\omega \sinh\left(\ln\frac{\omega}{c_1}+k_0r_c |\phi| \right)~~~~~\mathrm{for~AdS~brane}
\label{Eq2}
\end{align}  
where, $\omega=(\pm\Omega/3k_0^2)$ for dS and AdS case respectively, and $c_2=1+\sqrt{1+\omega ^2}$, $c_1=1+\sqrt{1-\omega ^2}$. 
Since the observed accelerated expansion of the universe can be explained by a positive brane cosmological constant, we 
will concentrate mainly on the warped braneworld scenario with de-Sitter branes. 

The presence of such non vanishing brane cosmological constant not only generalizes the RS scenario, but also self stabilizes the model by 
generating a stable radion potential in the four dimensional effective theory. This is unlike to the original RS scenario where the branes are flat, and thus, 
an adhoc bulk scalar field has to be considered by hand in order to stabilize the braneworld model. Considering the dynamical radion field 
as $T(x)$, the four dimensional effective action 
is given by \cite{Banerjee:2020uil},
\begin{align}
\mathcal{A}=\int d^4 x \sqrt{-g}\Bigg[\frac{R}{2 \kappa^2}
-\frac{1}{2}G(\xi)\partial^\mu\Phi\partial_\mu\Phi 
-8 M^3\kappa^2 V(\xi)\Bigg]~~,  
\label{Eq17} 
\end{align}
where $\kappa^2 = 8\pi G$ ($G$ is the four dimensional Newton's constant), $\Phi = fe^{-k_0T(x)\pi}$ (with $f = \sqrt{6M^3c_2^2/k_0}$) is the radion field 
having mass dimension [+1] and $\xi = \frac{\Phi}{f}$ is the dimensionless radion field. The potential and the 
non-canonical kinetic term for the radion field come with the following expressions \cite{Banerjee:2020uil}:
\begin{eqnarray}
V(\xi) = \frac{6\omega^2}{h(\xi)}~~~~\mathrm{and}~~~~
G(\xi)=\frac{\mathcal{G}(\xi)}{h(\xi)}+\frac{1}{c_2^2}\bigg[\frac{h^\prime(\xi)}{h(\xi)}\bigg]^2
\label{Eq18} 
\end{eqnarray}
respectively, where
\begin{eqnarray}
h\left(\xi\right)&=&\left\{\frac{c_2^2}{4}+\omega^2 \ln \xi+\frac{\omega^4}{4c_2^2}\left(\frac{1}{\xi^2}\right)-\frac{\omega^4}{4c_2^2}
-\frac{c_2^2}{4}\xi ^2 \right\}\nonumber\\
\mathcal{G}\left(\xi\right)&=&1+\frac{4}{3}\frac{\omega^2}{c_2^2}\left(\frac{1}{\xi^2}\right)\ln \xi -\frac{\omega^4}{c_2^4}\left(\frac{1}{\xi^4}\right)~~.
\label{Eq11}
\end{eqnarray}
Note that $V(\xi)$ is proportional to $\omega^2$, which indicates that the potential term for the radion field is solely generated due to the presence 
of a non-zero brane cosmological constant. The $V(\xi)$ has an inflection point at $\xi = \omega/c_2$ and the $G(\xi)$ experiences a zero crossing 
from positive to negative values. 
In Fig.[\ref{Fig_2a}] and Fig.[\ref{Fig_2b}] we plot the variation of $V$ and $G$ with the radion field $\xi$ for $\omega=10^{-3}$. 
Fig.[\ref{Fig_2b}] reveals that the non-canonical 
coupling to the kinetic term $G(\xi)$ exhibits a transition from a normal to a phantom regime (i.e from $G(\xi) > 0$ to $G(\xi) < 0$), 
where the phantom like behavior remains when $\xi$ lies in the range $\xi_i\leq \xi \leq\xi_f$, 
with $\xi_f$ denoting the zero crossing of $G(\xi)$. During the phantom regime the kinetic energy of the radion field gets negative, which in turn may 
violate the null energy condition of the radion field. Such violation of null energy condition is the essence for triggering a non-singular bounce 
during the early stage of the universe. 
This raises the question whether the radion field can be instrumental in giving rise to a bouncing universe which we address in the next section.

\begin{figure*}[t!]
\centering
\subfloat[\label{Fig_2a}]{\includegraphics[scale=0.60]{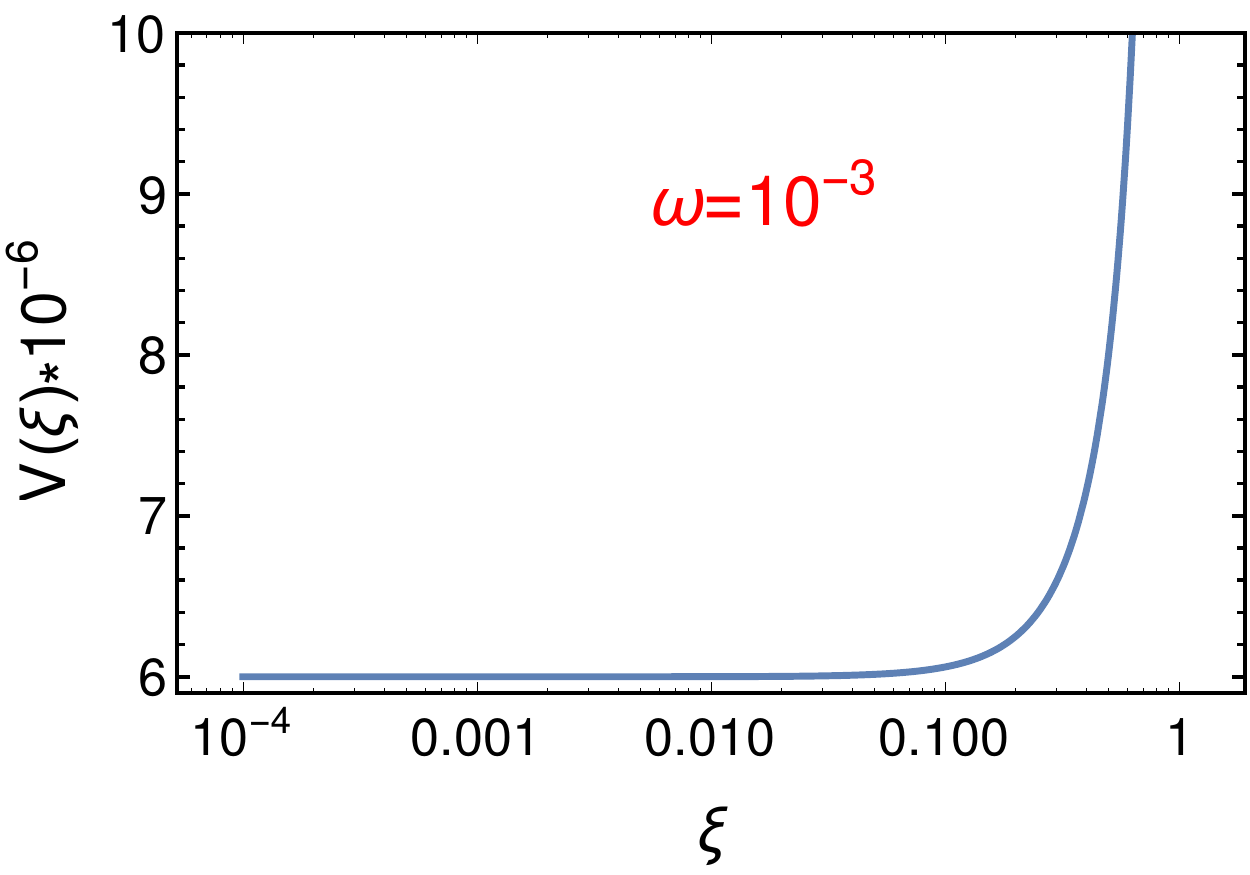}}~~
\subfloat[\label{Fig_2b}]{\includegraphics[scale=0.60]{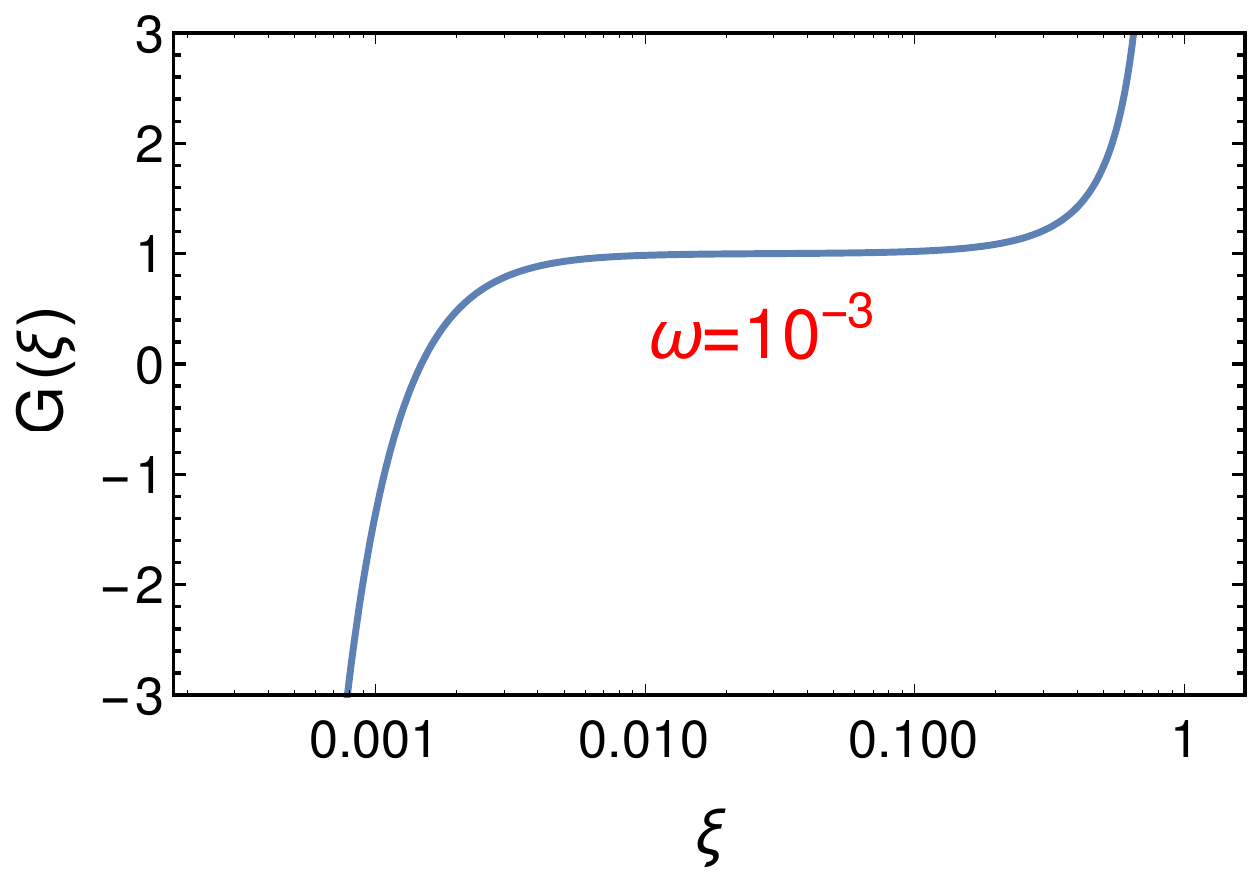}}
\caption{The above figure depicts the variation of (a) the radion potential $V$ and (b) the non-canonical 
coupling to the kinetic term $G$ in the Einstein frame, within the allowed range of the radion field $\xi$ for $\omega=10^{-3}$.}
\label{Fig_02}
\end{figure*}

\subsection{Implications in Early Universe Cosmology: Background evolution}
\label{S3}
The following metric ansatz will fulfill our purpose,
\begin{align}
ds^2=-dt^2+a(t)^2\bigg[dx^2+dy^2+dz^2\bigg] \label{Eq23} 
\end{align}
with $a(t)$ is known as the scale factor of the universe. Considering $\xi = \xi(t)$, the Friedmann equations for the action Eq.(\ref{Eq17}) are 
\cite{Banerjee:2020uil},
\begin{eqnarray}
H^2&=&\frac{\kappa^2}{3}\rho(t) =\frac{c_2^2}{4} G(\xi) \dot{\xi}^2 + \frac{k_0^2}{6} V(\xi)\nonumber\\
\dot{H}&=&-\frac{\kappa^2}{2}(\rho + p) = -\frac{3}{4}c_2^2 G(\xi) \dot{\xi}^2
\label{Eq28}
\end{eqnarray}
where $H=\dot{a}/a$ denotes the Hubble parameter. The equation of motion for the radion field is given by,
\begin{align}
\ddot{\xi} + 3H \dot{\xi} + \frac{G^\prime(\xi)}{2G(\xi)}\dot{\xi}^2 + \frac{k_0^2}{3c_2^2}\frac{V^\prime(\xi)}{G(\xi)}=0
\label{Eq30}
\end{align}
Eq.(\ref{Eq28}) reveals that the model has a possibility to show 
a bounce phenomena when the non-canonical function $G(\xi)$ becomes negative i.e when the radion field is in the phantom regime. 
$G(\xi)$ becomes negative in the regime $\xi \sim \omega$. Thus, at first we analytically 
solve the background equations near $\xi \sim \omega$ to investigate the bounce and then we numerically determine 
the background evolution for a wide range of $\xi$ (or equivalently for a wide range 
of cosmic time), where the boundary conditions of the numerical calculation are provided from the analytic solutions.\\
In particular we consider,
\begin{align}
\xi(t)=\frac{\omega}{c_2} [1+ \delta(t)]
\label{Eq31}
\end{align}
with $\delta(t) \ll 1$. In this regime of $\xi$, $V(\xi)$ and $G(\xi)$ are approximated by, 
\begin{eqnarray}
 V(\xi) \simeq \frac{24\omega^2}{c_2^2}~~~~~\mathrm{and}~~~~~
 G(\xi)\simeq -\frac{16}{3c_2^2}\bigg[ ln \bigg(\frac{c_2}{\omega}\bigg)-\bigg \lbrace 4 + 2 ln \bigg(\frac{c_2}{\omega}\bigg
)\bigg\rbrace \delta \bigg]~~.\nonumber
\end{eqnarray}
Using the above expressions of $V(\xi)$ and $G(\xi)$, Eq.(\ref{Eq28}) and Eq.(\ref{Eq30}) become
\begin{align}
\dot{H} + 3H^2 -\frac{12 k_0^2 \omega^2}{c_2^2}=0~~~~~\mathrm{and}~~~~~
\dot{\delta}^2=
\frac{c_2^2}{\omega^2}\frac{\dot{H}}{4 ln \big(\frac{c_2}{\omega}\big)}\bigg[1+  \delta \bigg\lbrace \frac{4+2ln\big(\frac{c_2}{\omega}\big)}{ln\big(\frac{c_2}{\omega}\big)}\bigg\rbrace \bigg]
\label{Eq36}
\end{align}
respectively. Solving which, we get the background solutions near $\xi \sim \omega$ as \cite{Banerjee:2020uil},
\begin{align}
\xi(t)=\frac{\omega}{c_2}(1+\delta(t)),
\begin{cases}
H(t)=2k_0\frac{\omega}{c_2}tanh \bigg[6 \frac{\omega}{c_2}k_0 t\bigg] & \\
    \delta(t)=\frac{2}{A}\bigg[exp \bigg\lbrace -\frac{A}{6}\frac{\omega}{c_2} \sqrt{\frac{3}{ln\big( \frac{c_2}{\omega} \big)}}\bigg ( tan^{-1} tanh \bigg(\frac{3\omega}{c_2}k_0 t\bigg) -\frac{\pi}{4}\bigg)   \bigg \rbrace -1  \bigg],              
\end{cases}
\label{Eq46}
\end{align}
where, $A=\frac{4+2ln \big(\frac{c_2}{\omega}\big)}{ln\big(\frac{c_2}{\omega}\big)}$. To derive the above solutions, we consider 
$\lim_{t\rightarrow \infty}\xi(t) = \frac{\omega}{c_2}$. Recall that $\xi = \frac{\omega}{c_2}$ is the inflection point of $V(\xi)$, and thus 
$\lim_{t\rightarrow \infty}\xi(t) = \frac{\omega}{c_2}$implies that the radion asymptotically reaches to its stable value. 

We now solve the coupled equations for $H(t)$ and $\xi(t)$ (i.e Eq.(\ref{Eq28})) for a wide range of cosmic time numerically. 
In regard to the numerical calculation, the boundary conditions are provided 
from the analytic solutions as determined in \ref{Eq46}, in particular, $H(0) = 0$ and 
$\xi(0) = 6.0041\times10^{-4}$, where we consider 
$\omega = 10^{-3}$ (later, during the perturbation calculation, we show that such a value of $\omega$ is consistent with 
the Planck 2018 constraints). The time evolution of the Hubble parameter and the radion field are shown in Fig.[\ref{Fig_3a}] and Fig.[\ref{Fig_3b}] 
respectively 
(more descriptions about the figure are given in the caption of the figure).
\begin{figure*}[t!]
\centering
\subfloat[\label{Fig_3a}]{\includegraphics[scale=0.60]{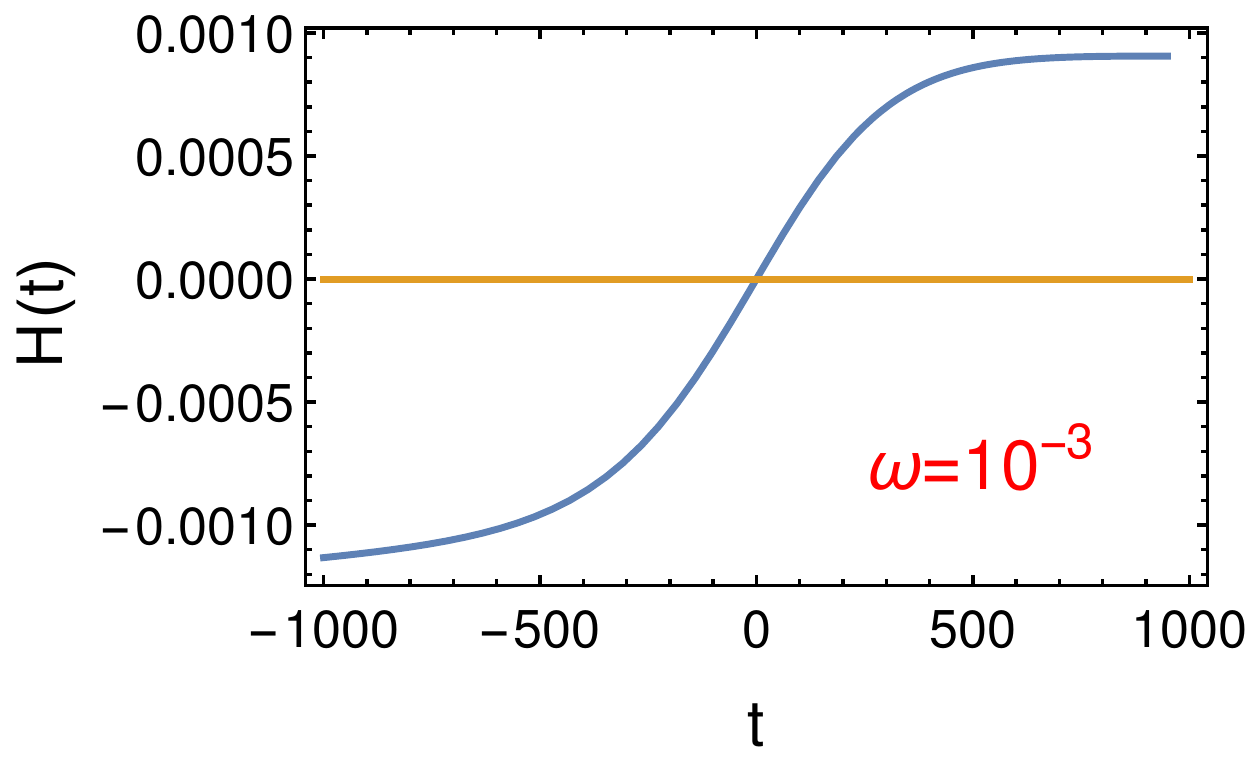}}~~
\subfloat[\label{Fig_3b}]{\includegraphics[scale=0.60]{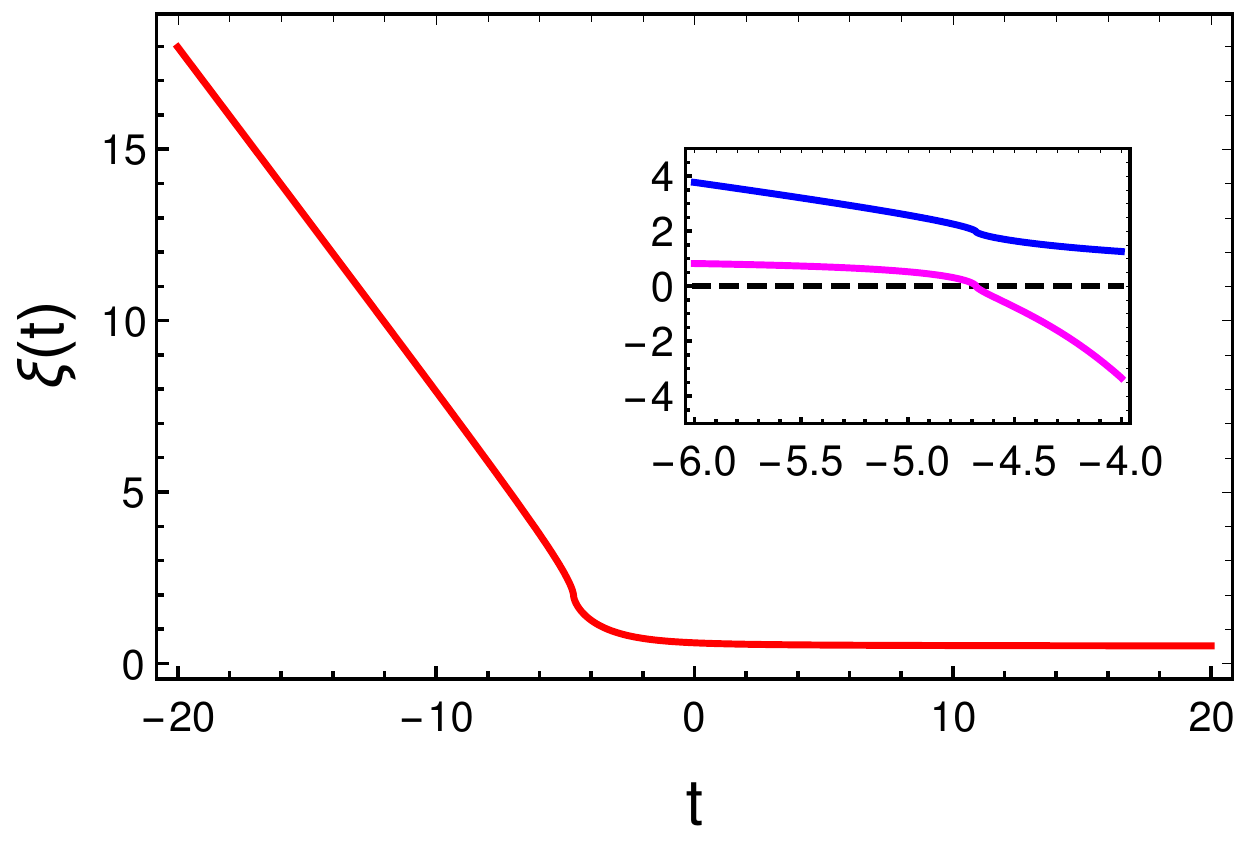}}
\caption{The above figure depicts the time evolution of (a) the Hubble parameter $H(t)$ and 
(b) the radion field magnified 1000 times, i.e. $\xi(t)\times1000$; while the inset of Fig.[\ref{Fig_3b}] 
depicts the non-canonical kinetic term $G(\xi)$ (magenta curve) 
and the zoomed-in version of $\xi(t)\times1000$ (blue curve) near the zero crossing of $G(\xi)$. 
Note that bounce occurs at $t=0$ when the kinetic term of the radion is in the phantom regime. 
Both the above figures are illustrated for $\omega=10^{-3}$.}
\label{Fig_03}
\end{figure*}

Fig.[\ref{Fig_03}] reveals -- (1) the Hubble parameter becomes zero and increases with respect to cosmic time at $t = 0$, 
which confirms a non-singular bounce 
at $t = 0$; (2) the radion field starts its journey from the normal regime 
and dynamically moves to the phantom era with time by monotonically decreasing in magnitude and asymptotically 
stabilizes to the value $\langle\xi\rangle \to \omega/c_2$ which for $\omega = 10^{-3}  \Rightarrow \langle\xi\rangle =5\times10^{-4}$.

\subsection{Implications in Early Universe Cosmology: Evolution of perturbations}
\label{S4}
Here we consider the spacetime perturbations over the background FRW metric and consequently determine the primordial 
observable quantities like the scalar spectral index ($n_s$), tensor to scalar ratio ($r$) and the amplitude of scalar perturbations ($A_s$). 
For the present bounce scenario, the comoving Hubble radius asymptotically goes to zero at both sides of the bounce, which in turn depicts that the 
perturbation modes generate near the bounce when the Hubble horizon is infinite in size to contain all the relevant modes within it. 
Therefore following we solve the perturbation equations near the bouncing point $t = 0$. In this regard we further mention that due to the reason that 
the comoving Hubble radius decreases (with time) at both sides of the bounce, the effective EoS parameter during the contracting stage is less 
than unity. In effect, the anisotropic energy density during the contracting phase grows faster than that of the bouncer, 
and leads to the BKL instability.

\begin{figure}[!h]
\begin{center}
 \centering
 \includegraphics[scale=1.00]{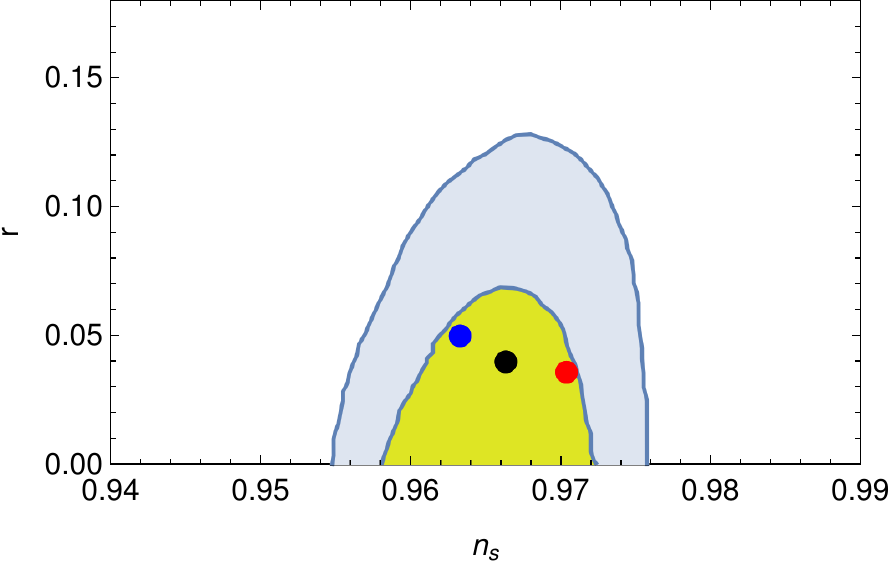}
 \caption{$1\sigma$ (yellow) and $2\sigma$ (light blue) contours for Planck 
 2018 results \cite{Akrami:2018odb}, on $n_s-r$ plane. 
 Additionally, we present the predictions of the present bounce scenario with $\frac{R_h}{\alpha} = 14.2$ (blue point), $\frac{R_h}{\alpha} = 16.2$ 
 (black point) and $\frac{R_h}{\alpha} = 18.8$ (red point).}
 \label{plot_observable}
\end{center}
\end{figure}

The scalar metric perturbation ($\Psi$) over the background FRW metric obeys the following equation in the longitudinal gauge, 
\begin{eqnarray}
 \ddot{\Psi} - \frac{1}{a^2}\nabla^2\Psi + \bigg[7H + \frac{2k_0^2~V'(\xi_0)}{3c_2^2G(\xi_0)\dot{\xi}_0}\bigg]\dot{\Psi} 
 + \bigg[2\dot{H} + 6H^2 + \frac{2k_0^2H~V'(\xi_0)}{3c_2^2G(\xi_0)\dot{\xi}_0}\bigg]\Psi = 0
 \label{sp8}
\end{eqnarray}
where $H = \frac{\dot{a}}{a}$ is the Hubble parameter in cosmic time and recall, $\xi_0$ 
is the dimensionless background radion field. Using the background solutions, 
we determine $\frac{V'(\xi_0)}{G(\xi_0)\dot{\xi}_0}$ (present in the above equation) near the bounce 
where we retain the term up-to the leading order in $t$:
\begin{eqnarray}
 \frac{V'(\xi_0)}{G(\xi_0)\dot{\xi}_0}
 &=&\frac{36\omega^2\delta^2(t)}{\dot{\delta}\big[\ln{\frac{\omega}{c_2}} + 2\big(2 - \ln{\frac{\omega}{c_2}}\big)\delta(t)\big]}\nonumber\\
 &=&-\frac{48B\omega c_2\sinh^2(B\pi/8)}{k_0\big(3 - 2e^{B\pi/4}\big)\big(2 - \ln{\frac{\omega}{c_2}}\big)} 
 + \frac{72\omega^2\big(2 - 2\cosh(B\pi/4) + \sinh(B\pi/4)\big)}{\big(3 - 2e^{B\pi/4}\big)^2\big(2 - \ln{\frac{\omega}{c_2}}\big)}~t
 \label{sp9}
\end{eqnarray}
with $B = \frac{A}{6}\frac{\omega}{c_2}\sqrt{\frac{3}{\ln{\big(\frac{c_2}{\omega}}\big)}}$. Consequently Eq.(\ref{sp8}) in Fourier space
becomes,
\begin{eqnarray}
 \ddot{\Psi}_k + \big[-\sqrt{\alpha}p + (q + 14)\alpha t\big]\dot{\Psi}_k + \big[k^2 + 4\alpha - 2\alpha\sqrt{\alpha}p~t\big]\Psi_k(t) = 0~~,
 \label{sp11}
\end{eqnarray}
where $\alpha=\frac{6k_0^2\omega^2}{c_2^2}$ and $p$ and $q$ have the following expressions,
\begin{eqnarray}
 p = 16\sqrt{\frac{2}{3}}\bigg(\frac{B\sinh^2(B\pi/8)}{\big(3 - 2e^{B\pi/4}\big)\big(2 - \ln{\frac{\omega}{c_2}}\big)}\bigg)~~~~~\mathrm{and}~~~~~
 q = \frac{8\big(2 - 2\cosh(B\pi/4) + \sinh(B\pi/4)\big)}{\big(3 - 2e^{B\pi/4}\big)^2\big(2 - \ln{\frac{\omega}{c_2}}\big)}
 \nonumber
\end{eqnarray}
respectively. Considering the Bunch-Davies initial condition of scalar Mukhanov-Sasaki variable $t = 0$, Eq.(\ref{sp11}) is solved to get,
\begin{eqnarray}
 \Psi_k(t) = \frac{\sqrt{3}}{2k^{3/2}}\bigg(\frac{\omega}{c_2}\bigg)\bigg(\frac{k_0}{M}\bigg)^{3/2}e^{B\pi/4}\big(3 - 2e^{B\pi/4}\big)^{1/2} 
 e^{[p\sqrt{\alpha}~t~ - 7\alpha t^2 - \frac{q}{2}\alpha t^2]}~\Bigg\{\frac{H\big[-1 + \frac{k^2 + 4\alpha}{\alpha(q+14)}, 
 \frac{-p + (q+14)\sqrt{\alpha}~t}{\sqrt{2(q+14)}}\big]}{H\big[-1 + \frac{k^2 + 4\alpha}{\alpha(q+14)}, \frac{-p}{\sqrt{2(q+14)}}\big]}\Bigg\}~~.
 \label{sp15}
\end{eqnarray}
The solution of $\Psi_k(t)$ immediately leads to the scalar power spectrum for 
$k$-th modes as,
\begin{eqnarray}
 P_{\Psi}(k,t) = 
 \frac{3}{8\pi^2}\bigg(\frac{\omega}{c_2}\bigg)^2\bigg(\frac{k_0}{M}\bigg)^{3}e^{B\pi/2}\big(3 - 2e^{B\pi/4}\big) 
 e^{[2p\sqrt{\alpha}~t~ - 14\alpha t^2 - q\alpha t^2]}~\Bigg\{\frac{H\big[-1 + \frac{k^2 + 4\alpha}{\alpha(q+14)}, 
 \frac{-p + (q+14)\sqrt{\alpha}~t}{\sqrt{2(q+14)}}\big]}{H\big[-1 + \frac{k^2 + 4\alpha}{\alpha(q+14)}, \frac{-p}{\sqrt{2(q+14)}}\big]}\Bigg\}^2~.
 \label{sp16}
\end{eqnarray}
The tensor perturbation near the bounce follows the equation like,
\begin{eqnarray}
 \ddot{h}_k + 6\alpha \dot{h}_k~t + k^2h_k(t) = 0
 \label{ten per eom3}
\end{eqnarray}
with recall, $\alpha=\frac{6k_0^2\omega^2}{c_2^2}$. Considering the Bunch-Davies state for tensor perturbation at $t = 0$, 
we solve Eq.(\ref{ten per eom3}) to get,
\begin{eqnarray}
 h_k(t) = \bigg(\frac{2\kappa~\Gamma\big(1 - \frac{k^2}{12\alpha}\big)}{\sqrt{2\pi k}~2^{\frac{k^2}{6\alpha}}}\bigg)~
 e^{-3\alpha t^2}~H\bigg[-1 + \frac{k^2}{6\alpha}, \sqrt{3\alpha}~t\bigg]~~,
 \label{ten per sol2}
\end{eqnarray}
which immediately
leads to the tensor power spectrum as,
\begin{eqnarray}
 P_{h}(k,t) = \frac{2k^2}{\pi^3}~\frac{\bigg(\kappa~\Gamma\big(1 - \frac{k^2}{12\alpha}\big)\bigg)^2}
 {~2^{\frac{k^2}{3\alpha}}} e^{-6\alpha t^2}~
 \bigg\{H\bigg[-1 + \frac{k^2}{6\alpha}, \sqrt{3\alpha}~t\bigg]\bigg\}^2
 \label{ten power spectrum}
\end{eqnarray}
Now one can explicitly confront the model at hand with the latest
Planck observational data \cite{Akrami:2018odb}, so we 
calculate the spectral index of the primordial curvature
perturbations $n_s$ and the tensor-to-scalar ratio $r$, which are defined and have the respective constraints from the Planck observation as follows,
\begin{eqnarray}
 n_s - 1&=&\frac{\partial\ln{P_{\Psi}}}{\partial\ln{k}}\bigg|_{h.c}~~,~~~~\mathrm{Constraint}:~~n_s = 0.9649 \pm 0.0042\nonumber\\
 r&=&\frac{P_h(k,t)}{P_{\Psi}(k,t)}\bigg|_{h.c}~~,~~~~\mathrm{Constraint}:~~r < 0.064\label{spectral index1}
\end{eqnarray}
where the suffix 'h.c' indicates the horizon crossing instance of the large scale modes. As mentioned earlier that the relevant perturbation modes 
cross the horizon near the bounce, and thus the horizon crossing condition becomes $k = aH = 2\alpha t_h$ (where $t_h$ is the horizon
crossing time). With this relation along with Eq.(\ref{spectral index1}), the parametric plot of $n_s$ vs. $r$ is shown in Fig.[\ref{plot_observable}] 
which clearly demonstrates the simultaneously compatibility of $n_s$ and $r$ with the Planck data. If one combines the scalar perturbation amplitude 
($A_s$) with $n_s$ and $r$, then $n_s$, $r$ and $A_s$ in the present bounce scenario become simultaneously compatible with the Planck 
constraints for the parameter ranges: $\omega = 10^{-3}$, $14 \lesssim \frac{R_h}{\alpha} \lesssim 19$, $\frac{k_0}{M} = [0.601 , 0.607]$; where 
$R_h$ is the Ricci scalar at the horizon crossing instant and recall that $k_0=\sqrt{-\Lambda/24M^3}$ 
such that $\Lambda$ and $M$ denote the five dimensional cosmological constant and Planck mass respectively.\\

To determine the scalar and tensor power spectra, we use the Hubble parameter as
\begin{eqnarray}
 H(t) = 12k_0^2\left(\frac{\omega}{c_2}\right)^2t~~~~~~~~~\mathrm{or~equivalently}~~~~~~~~~~a(t) = 1 + 6k_0^2\left(\frac{\omega}{c_2}\right)^2t^2~~,
 \label{revision-1}
\end{eqnarray}
where we keep up to the linear order in t from Eq.(\ref{Eq46}), and consequently, the evolution equations for scalar and tensor perturbations 
are considered up to the linear order in t. Therefore the scalar and tensor power spectra obtained in Eq.(\ref{sp16}) and Eq.(\ref{ten power spectrum}) 
respectively, are valid for 
\begin{eqnarray}
t < \frac{1}{k_0}\left(\frac{c_2}{6\omega}\right) = t_v (\mathrm{say})~~.
\label{revision-2}
\end{eqnarray}
As we have shown that the model stands to be a viable one in regard to the Planck constraints for the parameter ranges :
$\omega = 10^{-3}$  and $\frac{k_0}{M} = [0.601 , 0.607]$ respectively. Such parametric regime results to 
$t_v \sim 10^{-16}\mathrm{GeV}^{-1}$. On other hand, the horizon crossing condition for $k$-th mode is given by $k = aH$. 
Here we would like to mention that the scale of interest in the present context is around the CMB scale given by 
$k_{CMB} \approx 0.05\mathrm{Mpc}^{-1} \approx 10^{-40}\mathrm{GeV}$, as we are interested to investigate whether the theoretical predictions of 
$n_s$, $\mathcal{A}_s$ and $r$ match with the Planck 2018 results which put a constraint on these observable quantities 
around the CMB scale. With Eq.(\ref{revision-1}), we determine the expression of the time when $k_{CMB}$ 
crosses the horizon by using the horizon crossing relation $k = aH$, and is given by,
\begin{eqnarray}
 t_{h} = \frac{k_{CMB}}{12k_0^2}\bigg(\frac{c_2^2}{\omega^2}\bigg)~~,
 \label{new1}
\end{eqnarray}
where, $t_h$ is the horizon crossing time of the CMB scale and recall, $k_0$ being the bulk curvature scale. 
 Using the parametric regime mentioned above, we get 
the horizon crossing instance of $k_{CMB}$ as $t_{h} \sim 10^{-68}\mathrm{GeV}^{-1}$. This leads to $t_h \ll t_v$, which in turn justifies 
that the power spectra are evaluated when the large scale modes are on super-Hubble scales.

Thus as a whole, the presence of a non-vanishing brane cosmological constant results to a phantom phase of the radion field during its evolution. 
The existence of such phantom phase leads to a violation of null energy condition and triggers a non-singular bounce which predicts a low tensor-to scalar ratio and gets 
well consistent with the Planck data for suitable regime of the model parameters. 

\section{\underline{Ekpyrotic bounce driven by Kalb-Ramond field}}\label{sec-b2}
Here we review the ekpyrotic bounce scenario proposed in \cite{Paul:2022mup}, in particular, we 
explore the possible roles of Kalb-Ramond (KR) field in driving a non-singular bounce, 
in particular an ekpyrotic bounce driven by the second rank antisymmetric KR field in F(R) gravity theory. 
With a suitable conformal transformation 
of the metric, the F(R) frame can be mapped to a scalar-tensor theory, where the KR field gets coupled with the scalaron field (coming from higher 
curvature d.o.f) by a simple linear coupling. Such interaction between the KR and the scalaron field proves to be useful in violating 
the null energy condition and to trigger a non-singular bounce. In regard to the perturbation analysis, we examine the curvature power spectrum for 
two different scenario depending on the initial conditions: (1) in the first scenario, the universe initially 
undergoes through an ekpyrotic phase of contraction and consequently the large scales of primordial 
perturbation modes cross the horizon during the ekpyrotic stage, while, (2) in the second scenario, the ekpyrotic phase is preceded by 
a quasi-matter dominated pre-ekpyrotic stage, and thus the large scale modes (on which we are interested) cross the horizon during the pre-ekpyrotic phase. 
The existence of the pre-ekpyrotic stage seems to be useful in getting a nearly scale invariant curvature perturbation spectrum over the large 
scale modes. The detailed qualitative features are discussed.

\subsection{The model}
We start with a second rank antisymmetric Kalb-Ramond (KR) field in F(R) gravity, and the action is \cite{Paul:2022mup},
\begin{eqnarray}
 S = \int d^4x\sqrt{-g}\left[\left(\frac{1}{2\kappa^2}\right)F(R) - \frac{1}{12}H_{\mu\nu\alpha}H_{\rho\beta\delta}g^{\mu\rho}g^{\nu\beta}g^{\alpha\delta}
 \right]~,
 \label{jordan-action}
\end{eqnarray}
$\frac{1}{2\kappa^2}=M_\mathrm{Pl}^2$ ($M_\mathrm{Pl}$ being the Planck mass). 
$H_{\mu\nu\alpha}$ is the field strength tensor of KR field, defined by 
$H_{\mu\nu\alpha} = \partial_{[\mu}B_{\nu\alpha]}$. 
The above action can be mapped into the Einstein frame by using the following conformal transformation: 
$g_{\mu\nu} \longrightarrow \widetilde{g}_{\mu\nu} = \sqrt{\frac{2}{3}}\kappa\Phi(x^{\mu})~g_{\mu\nu}$, 
with $\Phi$ being the conformal factor and related to the spacetime curvature as $\Phi(R) = \frac{1}{\kappa}\sqrt{\frac{3}{2}}F'(R)$. 
Consequently the action in Eq.(\ref{jordan-action}) can be expressed as a scalar-tensor theory,
\begin{eqnarray}
 S = \int d^4x\sqrt{-\widetilde{g}}\left[\frac{\widetilde{R}}{2\kappa^2} - \frac{3}{4\kappa^2}\left(\frac{1}{\Phi^2}\right)
 \widetilde{g}^{\mu\nu}\partial_{\mu}\Phi\partial_{\nu}\Phi - V(\Phi) - \frac{1}{12}\left(\sqrt{\frac{2}{3}}\kappa\Phi\right)H_{\mu\nu\rho}H^{\mu\nu\rho}
 \right]~.
 \label{Einstein-action}
\end{eqnarray} 
where $\widetilde{R}$ is the Ricci scalar formed by $\widetilde{g}_{\mu\nu}$ and 
$H^{\mu\nu\rho} = H_{\alpha\beta\delta}\widetilde{g}^{\mu\alpha}\widetilde{g}^{\nu\beta}\widetilde{g}^{\rho\delta}$. The scalar field 
potential depends on the form of F(R), and is given by
\begin{eqnarray}
 V(\Phi) = \frac{1}{2\kappa^2}\left[\frac{RF'(R) - F(R)}{F'(R)^2}\right]~.
\end{eqnarray}
For our present purpose, we consider the isotropic and homogeneous FRW metric:
\begin{eqnarray}
 ds^2 = -dt^2 + a^2(t)\delta_\mathrm{ij}dx^\mathrm{i}dx^\mathrm{j}
 \label{FRW metric}
\end{eqnarray}
where $t$ and $a(t)$ are the cosmic time and the scale factor of the universe respectively. 
Here we would like to emphasize that $H_{\mu\nu\lambda}$ has four 
independent components in four dimensional spacetime due to its antisymmetric nature, and they can be expressed as,
\begin{eqnarray}
 H_\mathrm{012} = h_\mathrm{1}~~~,~~~H_\mathrm{013} = h_\mathrm{2}~~~,~~~H_\mathrm{023} = h_\mathrm{3}~~~,~~~H_\mathrm{123} = h_\mathrm{4}~.
 \label{independent components-KR}
\end{eqnarray}
However due to the isotropic and homogeneous spacetime, the off-diagonal Friedmann equations lead to the following solutions:
\begin{eqnarray}
 h_\mathrm{1} = h_\mathrm{2} = h_\mathrm{3} = 0~~~~~~~~\mathrm{and}~~~~~~~h_\mathrm{4} \neq 0~.
 \label{solution-off-diagonal-E-equation}
\end{eqnarray}
Using this solution, one easily obtains the independent field equations as follows,
\begin{eqnarray}
 H^2&=&\frac{\kappa^2}{3}
 \left[\frac{3}{4\kappa^2}\left(\frac{\dot{\Phi}}{\Phi}\right)^2 + V(\Phi) + \frac{1}{2}\sqrt{\frac{2}{3}}\kappa\Phi~h_\mathrm{4}h^\mathrm{4}\right]~,
 \label{diagonal-E-equations}\\
 \frac{\ddot{\Phi}}{\Phi}&-&\left(\frac{\dot{\Phi}}{\Phi}\right)^2 + 3H\left(\frac{\dot{\Phi}}{\Phi}\right) 
 + \left(\frac{2\kappa^2}{3}\right)\Phi~V'(\Phi) + \left(\frac{\sqrt{2}}{3\sqrt{3}}\right)\kappa^3\Phi~h_\mathrm{4}h^\mathrm{4} = 0
 \label{scalar field equation}\\
 0&=&\frac{1}{a^3}\partial_{\mu}\left(a^3\Phi H^{\mu\nu\lambda}\right)
 \label{KR field equation-1}
\end{eqnarray}
which are the Friedmann equation, the scalar field equation and the KR field equation respectively. 
A little bit of playing with the above equations lead to the following solutions of KR field energy density:
\begin{eqnarray}
 h_\mathrm{4}h^\mathrm{4} = h_\mathrm{0}/a^\mathrm{6}~,
 \label{solution-KR field}
\end{eqnarray}
with $h_0$ being an integration constant which is taken to be positive to ensure a real valued solution for $h^4(t)$. Furthermore, we consider 
an ansatz for the scalar field solution in terms of the scale factor as,
\begin{eqnarray}
 \Phi(t) = -\frac{1}{\kappa}\sqrt{\frac{3}{2}}\left(\frac{1}{a^\mathrm{n}(t)}\right)~,
 \label{ansatz-scalar field}
\end{eqnarray}
with $n > 0$. Eq.(\ref{ansatz-scalar field}) indicates that the scalar field acquires negative values during the cosmological evolution of the universe, 
which actually proves to be useful to get a non-singular bounce. With the above solutions for $\Phi(t)$ and $h_4(t)$, 
Eq.(\ref{diagonal-E-equations}) provides a two branch solution of $H = H(a)$:
\begin{eqnarray}
 H(a) = \pm a^\mathrm{-3n^2/4}\left\{C_\mathrm{1} + \frac{2\kappa^2h_\mathrm{0}a^\mathrm{\left(3n^2-2n-12\right)/2}}{\left(3n^2-2n-12\right)}\right\}^{1/2}~,
 \label{solution-Hubble parameter}
\end{eqnarray}
where $C_1$ is an integration constant. Therefore the evolution of $H(a)$ becomes different depending on whether $3n^2-2n-12 < 0$ or 
$3n^2-2n-12 > 0$. However both the cases will be proved to lead a non-singular bounce irrespective of the values of 
$n > 0$. Here we discuss the case when $3n^2-2n-12 < 0$ and its consequences, 
while the other case can be described by a similar fashion \cite{Paul:2022mup}.

Here we consider $3n^2 - 2n - 12 = -2q$, with $q > 0$, and consequently, the solution of $H(a)$ in Eq.(\ref{solution-Hubble parameter}) can be 
expressed as \cite{Paul:2022mup},
\begin{eqnarray}
 H(a) = \pm \sqrt{\frac{\kappa^2h_\mathrm{0}}{q}}~a^\mathrm{-3n^2/4}\left\{\frac{1}{a_\mathrm{0}^q} - \frac{1}{a^q}\right\}^{1/2}
 \label{solution-Hubble parameter-case-1}
\end{eqnarray}
where the integration constant $C_1$ is replaced by $a_0$ as $C_\mathrm{1} = \kappa^2h_\mathrm{0}/\left(qa_\mathrm{0}^q\right)$. 
The above expression of $H(a)$ satisfies the following two conditions at $a = a_0$,
\begin{eqnarray}
 H(a=a_0) = 0~~~~~~~~~~\mathrm{and}~~~~~~~~~~~\frac{dH}{dt}\bigg|_{a_0} = \left(aH\right)\frac{dH}{da}\bigg|_{a_0} = \frac{\kappa^2h_0}{2a_0^{6+n}}~,
 \label{bounce-confirm}
\end{eqnarray}
which clearly depicts that the universe experiences a non-singular bounce at $a=a_0$. 
Here it deserves mentioning that in absence of the KR field, the model is not able to 
predict a non-singular bounce of the universe. In particular, the Hubble parameter evolves as 
$H(a) \propto a^{-3n^2/4}$ when the KR field is absent, 
which does not lead to a bouncing scenario at all. It is important to realize that 
the KR field which has negligible footprints at present epoch of the universe, 
plays a significant role during the early universe to trigger a non-singular bounce.

\subsubsection*{Condition for $ekpyrotic$ character of the bounce}
During the deep contracting era, the Hubble parameter evolves as $H(a) \propto a^{-3n^2/4}$ (from Eq.(\ref{solution-Hubble parameter-case-1})), 
which immediately leads to the effective equation of state (EoS) parameter as,
\begin{eqnarray}
 \omega_\mathrm{eff} = -1 - \left(\frac{2a}{3H}\right)\frac{dH}{da} = -1 + \frac{n^2}{2}~.
 \label{eos-case-1}
\end{eqnarray}
Therefore in order to have an ekpyrotic character of the bounce, in which case $\omega_\mathrm{eff} > 1$, the parameter $n$ needs to satisfy $n>2$. 
Moreover the condition $3n^2-2n-12 < 0$ leads to $\frac{1}{3}\left(1-\sqrt{37}\right) < n < \frac{1}{3}\left(1+\sqrt{37}\right)$ which, along with 
$n>2$, provides,
\begin{eqnarray}
 2 < n < \frac{1}{3}\left(1 + \sqrt{37}\right)~.
 \label{constraint-1}
\end{eqnarray}
Since the bounce is ekpyrotic and symmetric, the energy density of the bouncing agent rapidly decreases with the expansion of the universe after the bounce 
(faster than that of the pressureless matter and radiation), and consequently the standard Big-Bang 
cosmology of the universe is recovered.

\subsection{Perturbation analysis}
Due to the ekpyrotic condition $n>2$, the comoving Hubble radius diverges to infinity at the deep contracting phase, which in turn indicates that 
the primordial perturbation modes generate during the contracting phase far away from the bounce when all the perturbation modes lie within the 
Hubble horizon. Moreover, since the model involves two fields, apart from the curvature perturbation, 
isocurvature perturbation also arises. In this regard, the ratio of the Kalb-Ramond to the scalaron field energy density is given by,
\begin{eqnarray}
 \frac{\rho_{KR}}{\rho_{\Phi}} = \left[\frac{\frac{1}{2}\sqrt{\frac{2}{3}}\kappa\Phi~h_\mathrm{4}h^\mathrm{4}}
 {\frac{3}{4\kappa^2}\left(\frac{\dot{\Phi}}{\Phi}\right)^2 + V(\Phi)}\right]~~.\nonumber
\end{eqnarray}
From the background solution of the KR and the scalaron field (that we have obtained earlier), one may argue that 
the Kalb-Ramond to the scalaron field energy density during the late stage of the universe goes by $1/a^{q}$ which tends to zero. This 
results to a weak coupling between the curvature and the isocurvature perturbations during the late evolution of the universe. 
However the ratio between $\rho_{KR}$ and $\rho_{\Phi}$ becomes comparable at the bounce, which in turn leads to a considerable coupling 
between the curvature and the isocurvature perturbations at the bounce. In the present context, we are interested 
to examine the curvature perturbation during the contracting universe (away from the bounce), and thus, 
we can safely ignore the coupling between the curvature and the isocurvature perturbations and solely concentrate on the curvature perturbation.

The Mukhanov-Sasaki (MS) variable of the curvature perturbation follows the equation (in the Fourier space and conformal time coordinate):
\begin{eqnarray}
 v_k''(\eta) + \left(k^2 - \frac{4\left(8-3n^2\right)}{\left(3n^2-4\right)^2\eta^2}\right)v_k(\eta) = 0~,
 \label{MS-equation-3}
\end{eqnarray}
on solving which for $v_k(\eta)$, we get,
\begin{eqnarray}
v(k,\eta) = \frac{\sqrt{\pi|\eta|}}{2}\left[c_1(k)H_{\nu}^{(1)}(k|\eta|) + c_2(k)H_{\nu}^{(2)}(k|\eta|)\right]\, ,
\label{MS-solution}
\end{eqnarray}
with $\nu = \sqrt{\frac{4\left(8-3n^2\right)}{\left(3n^2-4\right)^2} + \frac{1}{4}}$. 
Moreover $c_1$, $c_2$ are integration constants, $H_{\nu}^{(1)}(k|\eta|)$ and $H_{\nu}^{(2)}(k|\eta|)$ are the Hermite functions (having order 
$\nu$) of first and second kind, respectively. 
Considering the Bunch-Davies initial condition at the deep sub-Hubble scales, 
the curvature power spectrum in the super-Hubble scale becomes \cite{Paul:2022mup},
\begin{eqnarray}
\mathcal{P}(k,\eta) = \left[\left(\frac{1}{2\pi}\right)\frac{1}{z\left|\eta\right|}\frac{\Gamma(\nu)}{\Gamma(3/2)}\right]^2
\left(\frac{k|\eta|}{2}\right)^{3-2\nu}\, .
\label{curvature-power-spectrum-superhorizon}
\end{eqnarray}
Consequently the scalar spectral tilt becomes,
\begin{eqnarray}
 n_s = \frac{9n^2 - 4}{3n^2 - 4}~.
 \label{spectral-tilt-1}
\end{eqnarray}
As demonstrated in Eq.(\ref{constraint-1}), the parameter $n$ is constrained by $2 < n < \frac{1}{3}\left(1 + \sqrt{37}\right)$, 
which immediately leads to the following range of the spectral tilt: $3.6 < n_s < 4$. This indicates a blue-tilted curvature power spectrum, 
and thus is not consistent with the Planck 2018 results.

\begin{figure}[!h]
\begin{center}
\centering
\includegraphics[width=3.5in,height=3.0in]{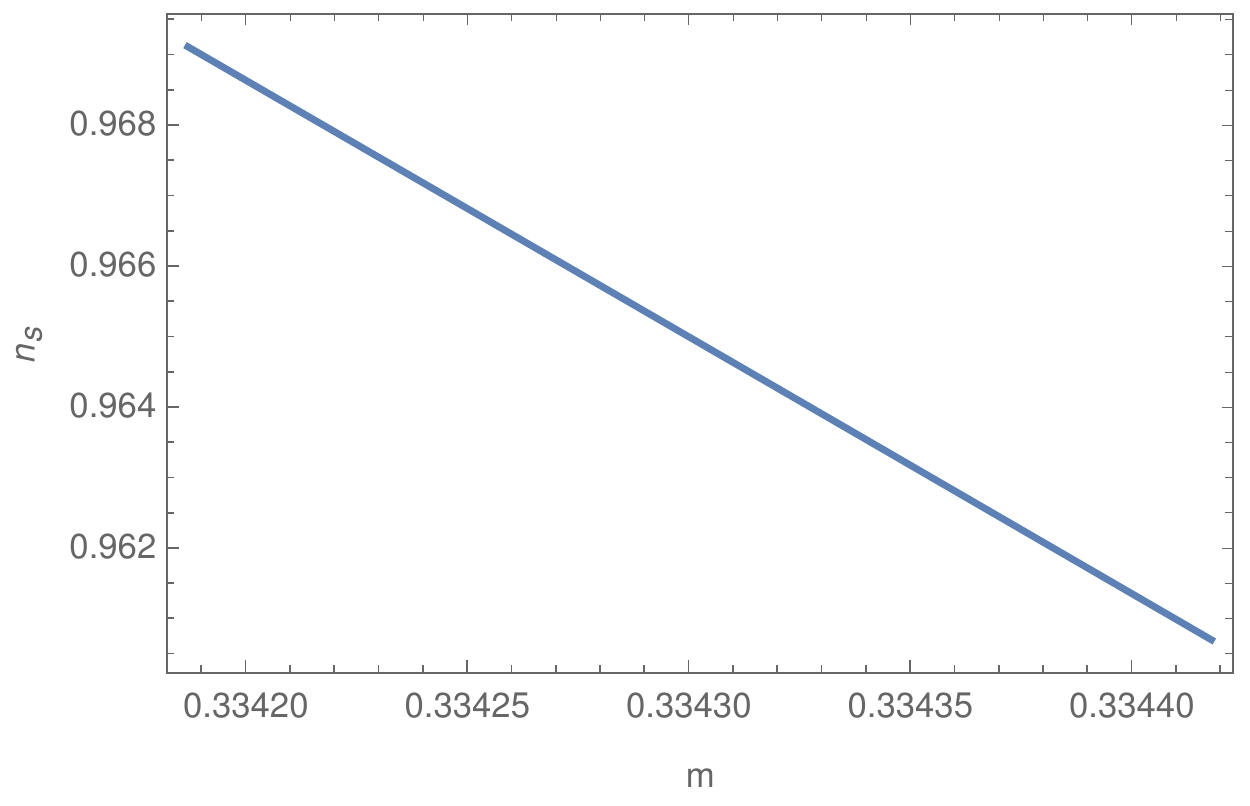}
\caption{$n_s$ vs. $m$ from Eq.(\ref{obs-1-pre})}
\label{plot-observable}
\end{center}
\end{figure}

In order to get a scale invariant curvature power spectrum in the present context, 
we consider a quasi-matter dominated pre-ekpyrotic phase where the scale factor behaves as $a_p(t) \sim t^{2m}$ with $m < 1/2$. Such a quasi-matter 
dominated phase can be realized by introducing a perfect fluid having constant EoS parameter $\approx 0$, in which case, the energy density 
grows as $\approx a^{-3}$ during the contracting phase. Thereby after some time, the KR field energy density (that grows as $a^{-{6+n}}$ with 
the universe's contraction, see Eq.(\ref{solution-KR field})) dominates over that of the perfect fluid 
and leads to an ekpyrotic phase of the universe.\\ 
In this modified cosmological scenario, the scale factor of the universe is:
\begin{align}
a_p(t) = a_\mathrm{1}\left(\frac{\eta - \eta_0}{\eta_e - \eta_0}\right)^{2m/\left(1-2m\right)} \quad &\mbox{with} \quad m<1/2\, ,  
\quad \mathrm{for}~\left|\eta\right| \geq \left|\eta_e\right|\, ,\nonumber\\
a(t) = a_2\left(-\eta\right)^{4/\left(3n^2 - 4\right)} \quad &\mbox{with} \quad 2<n<\frac{1}{3}\left(1+\sqrt{37}\right) \, , 
\quad \mathrm{for}~\left|\eta\right| \leq \left|\eta_e\right|\, .
\label{full-scale-factor}
\end{align} 
Here $\eta_e$ represents the conformal time when the transition from the pre-ekpyrotic to the 
ekpyrotic phase occurs, and $\eta_0$ is a fiducial time. Moreover the exponent $m < 1/2$ so that the comoving Hubble radius 
diverges at the distant past and the perturbation modes generate at the deep contracting phase within the sub-Hubble regime. 
As a whole, in this modified cosmological scenario, the scale factor of the universe is:
\begin{align}
a_p(t) = a_\mathrm{1}\left(\frac{\eta - \eta_0}{\eta_e - \eta_0}\right)^{2m/\left(1-2m\right)} \quad &\mbox{with} \quad m<1/2\, ,  
\quad \mathrm{for}~\left|\eta\right| \geq \left|\eta_e\right|\, ,\nonumber\\
a(t) = a_2\left(-\eta\right)^{4/\left(3n^2 - 4\right)} \quad &\mbox{with} \quad 2<n<\frac{1}{3}\left(1+\sqrt{37}\right) \, , 
\quad \mathrm{for}~\left|\eta\right| \leq \left|\eta_e\right|\, .
\label{full-scale-factor}
\end{align} 
The continuity of the scale factor as well as of the Hubble parameter at the transition time $\eta=\eta_e$ result to the following expressions, 
\begin{eqnarray}
a_1&=&a_2\left(-\eta_e\right)^{4/\left(3n^2 - 4\right)}~,\nonumber\\
\left(\frac{2m}{1-2m}\right)\frac{1}{\left(\eta_e - \eta_0\right)}&=&\left(\frac{4}{3n^2 - 4}\right)\frac{1}{\eta_e}~,
\label{continuity}
\end{eqnarray}
respectively. In effect of the pre-ekpyrotic phase of contraction, the large scale perturbation modes cross the horizon either during the pre-ekpyrotic 
or during the ekpyrotic stage depending on whether the transition time ($\eta_e$) is larger than the horizon crossing instant of the large scale modes 
($\eta_h$) or not. For a scale invariant power spectrum, here 
we consider the first case, i.e when the large scale modes cross the horizon during the pre-ekpyrotic phase. 
Thus the horizon crossing instant for $k$th mode is given by,
\begin{eqnarray}
\left|\eta_h\right| = \left(\frac{2m}{1-2m}\right)\frac{1}{k}\, .
\label{hc-1-pre-ekpyrotic}
\end{eqnarray}
Consequently the horizon crossing instant for the large scale modes, in particular $k = 0.002\mathrm{Mpc}^{-1}$, is estimated as,
\begin{eqnarray}
\left|\eta_h\right| \approx \left(\frac{2m}{1-2m}\right)\times13\,\mathrm{By}\, .
\label{hc-2-pre-ekpyrotic}
\end{eqnarray}
Thus one may argue that the transition from the pre-ekpyrotic to the ekpyrotic phase occurs at $\left|\eta_e\right| < 13\mathrm{By}$ so that 
the large scale modes cross the horizon during the pre-ekpyrotic era. Following 
the same procedure as of the previous section, we calculate the spectral tilt for the curvature perturbation in the modified 
scenario where the ekpyrotic phase is preceded by a period of a pre-ekpyrotic contraction:
\begin{eqnarray}
n_s = \frac{5- 14m}{1 - 2m}\, .
\label{obs-1-pre}
\end{eqnarray}
Clearly for $m = 1/3$ which describes a matter dominated epoch before the ekpyrotic phase, the spectral tilt becomes unity -- i.e an exact scale invariant 
power spectrum is predicted when the curvature perturbations over the large scale modes generate during a matter dominated pre-ekpyrotic era. 
However the observations according to the Planck data depict that the curvature power spectrum should not be exactly flat, but a has a slight red tilt. 
For this purpose, we give a plot of $n_s$ with respect to $m$ in Fig.[\ref{plot-observable}]. The figure clearly 
demonstrates that the theoretical prediction of $n_s$ becomes consistent with the Planck 2018 data if the parameter $m$ lies 
within $0.3341 \lesssim m \lesssim 0.3344$. Therefore the spectral index for the primordial curvature perturbation, on scales that 
cross the horizon during the pre-ekpyrotic stage with $0.3341 \lesssim m \lesssim 0.3344$, is found to be consistent with the recent Planck 
observations.\\

Before concluding, here we would to like mention that in addition to the scalar type perturbation, primordial tensor perturbation is 
also generated in the deep contracting phase from the Bunch-Davies state. 
The recent Planck data puts an upper bound on the tensor perturbation amplitude, in particular on the 
tensor to scalar ratio as $< 0.064$. However the Mukhanov-Sasaki equation for the 
tensor perturbation in the present context becomes analogous compared to that of the scalar perturbation, and thus 
both type of perturbations evolve in a similar way. This makes the tensor to scalar ratio in the 
present bounce model too large to be consistent with the Planck observation. There are some ways to circumvent this problem, like - 
(1) by amplifying the curvature perturbation from the gradient instability of 
$c_\mathrm{s}^2$ (sound speed) changing sign during the bounce, (3) by introducing Gauss-Bonnet higher curvature terms in the action \cite{Nojiri:2022xdo} 
etc. This will be an interesting generalization of the present scenario by introducing such mechanisms that may 
reduce the tensor to scalar ratio. We leave this particular topic for future study.

\section{\underline{Smooth unification from a bounce to the dark energy era}}\label{sec-b3}

\subsection{The model}\label{sec-model}

Here we consider F(R) gravity theory with Chern-Simons generalization. 
The gravitational action is given by,
\begin{eqnarray}
 S = \int d^4x\sqrt{-g}\frac{1}{2\kappa^2}\left[F(R) + \frac{1}{8}\nu(R)~\tilde{R}^{\mu\nu\alpha\beta}R_{\mu\nu\alpha\beta}\right]
 \label{action}
\end{eqnarray}
where $\nu(R)$ is known as Chern-Simons coupling function, 
$\tilde{R}^{\mu\nu\alpha\beta} = \epsilon^{\gamma\delta\mu\nu}R_{\gamma\delta}^{~~\alpha\beta}$, 
$\kappa^2$ stands for $\kappa^2 = 8\pi G = \frac{1}{M_\mathrm{Pl}^2}$ and also $M_\mathrm{Pl}$ 
is the reduced Planck mass. By using the metric formalism, we vary the
action with respect to the metric tensor $g_{\mu\nu}$, and the gravitational equations read,
\begin{eqnarray}
 F'(R)R_{\mu\nu} - \frac{1}{2}F(R)g_{\mu\nu} - \nabla_{\mu}\nabla_{\nu}F'(R) + g_{\mu\nu}\Box F'(R) = T^{(c)}_{\mu\nu}~~,
 \label{basic2}
\end{eqnarray}
with
\begin{eqnarray}
 T^{(c)}_{\mu\nu}&=&
 \frac{2}{\sqrt{-g}}\frac{\delta}{\delta g^{\mu\nu}}\left\{\frac{1}{8}\sqrt{-g}~\nu(R)\tilde{R}^{\mu\nu\alpha\beta}R_{\mu\nu\alpha\beta}\right\}\nonumber\\
 &=&\nu'(R)\tilde{R}^{\mu\nu\alpha\beta}R_{\mu\nu\alpha\beta}\left(\frac{\delta R}{\delta g^{\mu\nu}}\right) 
 + \epsilon_{\mu}^{~cde}\left[\nu_{,e;f}R^{f}_{~\nu cd} - 2\nu_{,e}R_{\nu c;d}\right]
 \label{EM tensor-CS}
\end{eqnarray}
is the energy-momentum tensor contributed from the Chern-Simons term \cite{Hwang:2005hb}. Moreover 
$R_{\mu\nu}$ is the Ricci tensor constructed from $g_{\mu\nu}$ and $\nu'(R) = \frac{d\nu}{dR}$. 
The background metric of the Universe will be assumed to be a flat Friedmann-Robertson-Walker (FRW) metric,
\begin{eqnarray}
 ds^2 = -dt^2 + a^2(t)\big[dx^2 + dy^2 + dz^2\big]
 \label{basic3}
\end{eqnarray}
with $a(t)$ being the scale factor of the Universe. 
The energy-momentum tensor $T^{(c)}_{\mu\nu}$ identically vanishes in the background of FRW spacetime, i.e we may argue that 
the Chern-Simons term does not affect the background Friedmann equations, as also stressed in \cite{Hwang:2005hb}. However as we will see later that the 
Chern-Simons term indeed affects the perturbation evolution over the FRW spacetime, particularly the tenor type perturbation. Hence 
the temporal and spatial components of Eq.(\ref{basic2}) become,
\begin{eqnarray}
0&=&-\frac{F(R)}{2} + 3\big(H^2 + \dot{H}\big)F'(R) - 18\big(4H^2\dot{H} + H\ddot{H}\big)F''(R)\nonumber\\
0&=&\frac{F(R)}{2} - \big(3H^2 + \dot{H}\big)F'(R) + 6\big(8H^2\dot{H} + 4\dot{H}^2 + 6H\ddot{H}
+ \dddot{H}\big)F''(R) + 36\big(4H\dot{H} + \ddot{H}\big)^2F'''(R)~~,
\label{basic4}
\end{eqnarray}
where $H = \dot{a}/a$ denotes the Hubble parameter of the 
Universe. These are the basic equations for background evolution, which we will use in the subsequent sections.

\subsection{Background evolution}\label{sec-background}
Here we are interested in getting a smooth unified cosmological picture from a non-singular bounce to the late time 
dark energy epoch. In this regard, the background scale factor is taken as \cite{Odintsov:2021yva},
\begin{eqnarray}
 a(t) = \left[1 + a_0\left(\frac{t}{t_0}\right)^2\right]^n\exp{\left[\frac{1}{(\alpha-1)}\left(\frac{t_s - t}{t_0}\right)^{1-\alpha}\right]} 
 = a_1(t) \times a_2(t) (\mathrm{say})~,
 \label{scale factor1}
\end{eqnarray}
where $a_0$, $n$, $\alpha$ are positive valued dimensionless parameters, 
while the other ones like $t_s$ and $t_0$ have the dimensions of time. The parameter 
$t_0$ is taken to scale the cosmic time in billion years, so we take $t_0 = 1\mathrm{By}$ (the $\mathrm{By}$ stands for 'billion years' 
throughout the paper). The scale factor is taken as a product of two factors- $a_1(t)$ and $a_2(t)$ respectively, where the factor $a_2(t)$ is motivated 
in getting a viable dark energy epoch at late time. Actually 
$a(t) = a_1(t)$ is sufficient for getting a non-singular bouncing universe where the bounce occurs at $t = 0$. However 
the scale factor $a_1(t)$ alone does not lead to a viable dark energy model 
according to the Planck results. Thereby, in order to get a bounce along with a viable dark energy epoch, we consider the scale factor 
as of Eq.(\ref{scale factor1}) where $a_1(t)$ is multiplied by $a_2(t)$. 
We will show that the presence of $a_2(t)$ does not harm the bouncing character of the universe, however it slightly shifts the bouncing time 
from $t = 0$ to a negative time and moreover the above scale factor leads to an asymmetric bounce scenario (as $a(t) \neq a(-t)$). 

The Hubble parameter and the Ricci scalar from the scale factor of Eq.(\ref{scale factor1}) turn out to be,
\begin{eqnarray}
 H(t) = \frac{1}{a}\frac{da}{dt} = \frac{2a_0nt}{\left(1 + a_0t^2\right)} + \frac{1}{\left(t_s - t\right)^{\alpha}}
 \label{Hubble parameter}
\end{eqnarray}
and
\begin{eqnarray}
 R(t) = \frac{12a_0n}{\left(1 + a_0t^2\right)^2}\left\{1 - a_0t^2\left(1-4n\right)\right\} 
 + \frac{12}{\left(t_s - t\right)^{2\alpha}} + \frac{6\alpha}{\left(t_s - t\right)^{1+\alpha}} + \frac{48a_0nt}{\left(1 + a_0t^2\right)
 \left(t_s - t\right)^{\alpha}}
 \label{ricci scalar}
\end{eqnarray}
respectively. Eq.(\ref{Hubble parameter}) refers different types of finite time singularity at $t = t_s$, 
in particular -- (1) the singularity is a Type-I singularity for $\alpha > 1$, (2) 
for $0<\alpha<1$, a Type-III singularity appears at $t = t_s$, (3) $-1<\alpha<0$ refers a Type-II singularity and (4) a Type-IV singularity 
arises for $\alpha < -1$ and non-integer. 
Therefore the finite time singularity at $t = t_s$ is almost inevitable in the present context. Thus in order to describe a singularity free universe's 
evolution up-to the present epoch ($\approx 13.5\mathrm{By}$), we consider the parameter $t_s$ to be greater than the present age of the universe, 
i.e $t_s > t_p \approx 13.5\mathrm{By}$. Therefore with this condition, we may argue that the Hubble parameter of Eq.(\ref{Hubble parameter}) 
describes a singularity free cosmological evolution up-to $t \gtrsim t_p$. During the cosmic time $t \gg t_p$ : either the universe will hit to the 
finite time singularity at $t = t_s$ (predicted by the present model) 
or possibly some more fundamental theory will govern that regime by which the finite time singularity can be avoided. 

In regard to the background evolution at late contracting era when the primordial perturbation modes generate within the deep 
sub-Hubble radius -- the scale factor, Hubble parameter and the Ricci scalar 
have the following expressions:
\begin{eqnarray}
 a(t) \approx a_0^nt^{2n}~~~~~~~~,~~~~~~~~H(t) \approx \frac{2n}{t}~~~~~~~~\mathrm{and}~~~~~~~~R(t) \approx -\frac{12n(1-4n)}{t^2}~~.
 \label{late contracting expressions}
\end{eqnarray}
With these expressions, the F(R) gravitational Eq.(\ref{basic4}) turns out to be, 
\begin{eqnarray}
 \left(\frac{2}{1 - 4n}\right)R^2\frac{d^2F}{dR^2} - \left(\frac{1 - 2n}{1 - 4n}\right)R\frac{dF}{dR} + F(R) = 0~~,
 \label{gravitational equation-late contracting}
\end{eqnarray}
on solving which, we get the form of F(R) at late contracting era as,
\begin{eqnarray}
 F(R) = R_0\left[\left(\frac{R}{R_0}\right)^{\rho} + \left(\frac{R}{R_0}\right)^{\delta}\right]
 \label{FR solution-late contracting}
\end{eqnarray}
where $R_0$ is a constant, and the exponents $\rho$, $\delta$ have the following forms (in terms of $n$),
\begin{eqnarray}
 \rho = \frac{1}{4}\left[3 - 2n - \sqrt{1 + 4n\left(5+n\right)}\right]~~~~~,~~~~~
 \delta = \frac{1}{4}\left[3 - 2n + \sqrt{1 + 4n\left(5+n\right)}\right]
 \label{rho-delta}
\end{eqnarray}
respectively. The above expression will be useful in determining the evolution of scalar and tensor perturbations. In the context of 
Chern-Simons F(R) gravity, the condition $F'(R)>0$ indicates the stability for both the curvature and the tensor perturbation. 
Eq.(\ref{FR solution-late contracting}) depicts that $F'(R)$ is positive for $n < 1/4$, and thus we take $n < 1/4$ in order to make 
the perturbations stable. 

\subsubsection*{Realization of a non-singular asymmetric bounce}\label{sec-bounce-realization}
In this section, we will show that the scale factor of Eq.(\ref{scale factor1}) allows a non-singular bounce at a finite negative time. 
As the parameters $a_0$, $n$ and $\alpha$ are positive, the Hubble parameter during $t > 0$ remains positive. However during negative time, 
i.e for $t < 0$, the first term of Eq.(\ref{Hubble parameter}) becomes negative while the second term remains positive, 
thus there is a possibility to have $H(t) = 0$ and $\dot{H} > 0$ at some negative $t$. Let us check it more explicitly. 
For $t < 0$, we can write $t = -|t|$ and Eq.(\ref{Hubble parameter}) can be expressed as,
\begin{eqnarray}
 H(t) = -\frac{2a_0n|t|}{\left(1 + a_0|t|^2\right)} + \frac{1}{\left(t_s + |t|\right)^{\alpha}} = -H_1(t) + H_2(t)~(\mathrm{say})~~.
 \label{Hubble parameter-bounce2}
\end{eqnarray}
The term $H_1(t)$ starts from the value zero at $t \rightarrow -\infty$ and reaches to zero at $t = 0$, with an extremum (in particular, a maximum) 
at an intermediate stage of $-\infty < t < 0$. However the second term $H_2(t)$ starts from the value zero at $t \rightarrow -\infty$ and reaches 
to $1/t_s^{\alpha}$ at $t = 0$, with a monotonic increasing behaviour during $-\infty < t < 0$. 
Furthermore both the $H_1(t)$ and $H_2(t)$ increase at 
$t \rightarrow -\infty$ and $H_1(t)$ increases at a faster rate compared to that of $H_2(t)$ for $\alpha > 1$. 
Here it may be mentioned that the condition $\alpha > 1$ is also related to the positivity of the Ricci scalar, as 
we will establish in Eq.(\ref{model parameter constraint-1}), and thus $\alpha > 1$ is well justified in the present context. 
These arguments clearly indicate that there 
exits a negative finite $t$, say $t = -\tau$ (with $\tau$ being positive), when the conditions for non-singular bounce holds. The 
bounce instant can be determined by the condition $H(-\tau) = 0$, i.e,
\begin{eqnarray}
 \frac{2a_0n\tau}{\left(1 + a_0\tau^2\right)} = \frac{1}{\left(t_s + \tau\right)^{\alpha}}~~.
 \label{bounce-instant}
\end{eqnarray}
In regard to the time evolution of the Ricci scalar, Eq.(\ref{late contracting expressions}) clearly indicates that $R(t)$ behaves as $\sim -1/t^2$ 
at distant past, i.e the Ricci scalar starts from $0^{-}$ at $t \rightarrow -\infty$. However at the instant of bounce, the $R(t)$ becomes positive, 
due to the reason that the Hubble parameter vanishes and its derivative is positive at the bounce point. Therefore the Ricci scalar 
must undergo a zero crossing from negative to positive value before the bounce occurs. At this stage, 
we require that after that zero crossing, the Ricci scalar 
remains to be positive throughout the cosmic time, which can be realized by a more stronger condition that the Ricci scalar has to be positive 
during the expanding phase of the universe, in particular,
\begin{eqnarray}
 R(t > -\tau) > 0~~,
 \label{requirement1}
\end{eqnarray}
which leads to the following relations between the model parameters \cite{Odintsov:2021yva}:
\begin{eqnarray}
 \alpha > 1~~~~~~~~\mathrm{and}~~~~~~~
 \frac{\sqrt{a_0}\left(1 - 4n\right)}{8\sqrt{3}} < \frac{1}{t_s^{\alpha}} < \frac{\sqrt{a_0}\left(1 - 4n\right)}{4\sqrt{3}}
 \label{model parameter constraint-1}
\end{eqnarray}
respectively. One of the above constraints $\alpha > 1$ leads to a Type-I singularity at $t = t_s$. 
However, since $t_s \gtrsim t_p$, the 
present model satisfactorily describes a singularity free cosmological evolution up-to $t \gtrsim t_p$ with $t_p \approx 13.5\mathrm{By}$ 
being the present age of the universe.

\subsubsection*{Acceleration and deceleration stages of the expanding universe}

The acceleration factor of the universe is given by $\ddot{a}/a = \dot{H} + H^2$ which, from Eq.(\ref{Hubble parameter}), turns out the be,
\begin{eqnarray}
 \frac{\ddot{a}}{a} = \frac{2a_0n\left\{1 - a_0t^2(1 - 2n)\right\}}{\left(1 + a_0t^2\right)^2} 
 + \frac{\alpha}{\left(t_s - t\right)^{1+\alpha}} + \frac{4a_0nt}{\left(1 + a_0t^2\right)\left(t_s - t\right)^{\alpha}} 
 + \frac{1}{\left(t_s - t\right)^{2\alpha}}~~.
 \label{acceleration-1}
\end{eqnarray}
It is evident that near $t \approx 0$, $\frac{\ddot{a}}{a} \approx 2a_0n + \frac{\alpha}{t_s^{1+\alpha}} + \frac{1}{t_s^{2\alpha}}$, 
i.e $\ddot{a}$ is positive. 
This is however expected, because $t \approx 0$ is the bouncing regime where, 
due to the fact that $\dot{H} > 0$ near the bounce, the universe undergoes through an accelerating stage. However, as $t$ increases particularly 
during $t^2 > \frac{1}{a_0(1 - 2n)}$, the first term of Eq.(\ref{acceleration-1}) becomes negative and hence the universe may expand through 
a decelerating phase. As $t$ increases further, the terms containing $1/(t_s - t)$ starts to grow at a faster rate compared to the other terms 
(since $\alpha$ is positive) and $\ddot{a}$ becomes positive, i.e the universe transits from the intermediate decelerating phase to an accelerating one. 
The first transition from the early acceleration (near the bounce) to a deceleration occurs at
\begin{eqnarray}
 t \simeq \frac{1}{\sqrt{a_0\left(1-2n\right)}}\left(1 + \frac{a_0}{t_s^{\alpha}\left(a_0(1-2n)\right)^{3/2}}\right) = t_1 (\mathrm{say})~,\label{t1}
\end{eqnarray}
while the second transition from the intermediate deceleration to an accelerating phase happens at,
\begin{eqnarray}
 t \simeq \frac{t_s}{2\alpha} = t_2 (\mathrm{say})~~.\label{t2}
\end{eqnarray}
The second transition from $\ddot{a} < 0$ to $\ddot{a} > 0$ 
is identified with the $late~time~acceleration$ epoch of the universe. Therefore we require $t_2 \lesssim t_p$, where $t_2$ is the instant 
of the second transition and recall, $t_p$ represents the present age of the universe.

The EoS parameter of the dark energy epoch is defined as $\omega_\mathrm{eff}(t) = -1-\frac{2\dot{H}}{3H^2}$, where $H(t)$ is shown 
in Eq.(\ref{Hubble parameter}). With this expression of $\omega_\mathrm{eff}$, we confront the model with the latest Planck+SNe+BAO results which put 
a constraint on the dark energy EoS parameter as \cite{Aghanim:2018eyx},
\begin{eqnarray}
 \omega_\mathrm{eff}(t_p) = -0.957 \pm 0.080
 \label{Planck1}
\end{eqnarray}
with $t_p \approx 13.5\mathrm{By}$ being the present age of the universe. Thereby we choose the model parameters in such a way that the above constraint 
on $\omega_\mathrm{eff}(t_p)$ holds true.\\

As a whole, we have four parameters in our hand: $n$, $a_0$, $t_s$ and $\alpha$. Below is the list of their constraints that we found earlier from 
various requirements,
\begin{itemize}
 \item $C1$ : The parameter $n$ is constrained by $n < 1/4$ in order to make the primordial perturbations stable at the deep sub-Hubble radius 
 in the contracting era.
 
 \item $C2$ : $t_s$ is larger than the present age of the universe, i.e $t_s > t_p \approx 13.5\mathrm{By}$ to describe a singularity free evolution 
 of the universe up-to the cosmic time $t \gtrsim t_p$.
 
 \item $C3$ : $t_s \lesssim 2\alpha t_p$ in order to have an accelerating stage of the present universe. This along with the previous condition 
 $C2$ lead to $t_p < t_s \lesssim 2\alpha t_p$.
 
 \item $C4$ : In regard to the parameters $\alpha$ and $a_0$, they are found to be constrained as $\alpha > 1$ and 
 $\frac{\sqrt{a_0}\left(1 - 4n\right)}{8\sqrt{3}} < \frac{1}{t_s^{\alpha}} < \frac{\sqrt{a_0}\left(1 - 4n\right)}{4\sqrt{3}}$. These make 
 the Ricci scalar positive after its zero crossing at the contracting era. In particular, the zero crossing (from negative to positive values) of the 
 Ricci scalar occurs before the instant of the bounce.
 
 \item $C5$ : $\omega_\mathrm{eff}(t_p) = -0.957 \pm 0.080$, to confront the theoretical expectations of the dark energy EoS with the Planck+SNe+BAO 
 results.
 \end{itemize}

 \begin{figure*}[t!]
\centering
\subfloat[\label{Fig_2a}]{\includegraphics[scale=0.35]{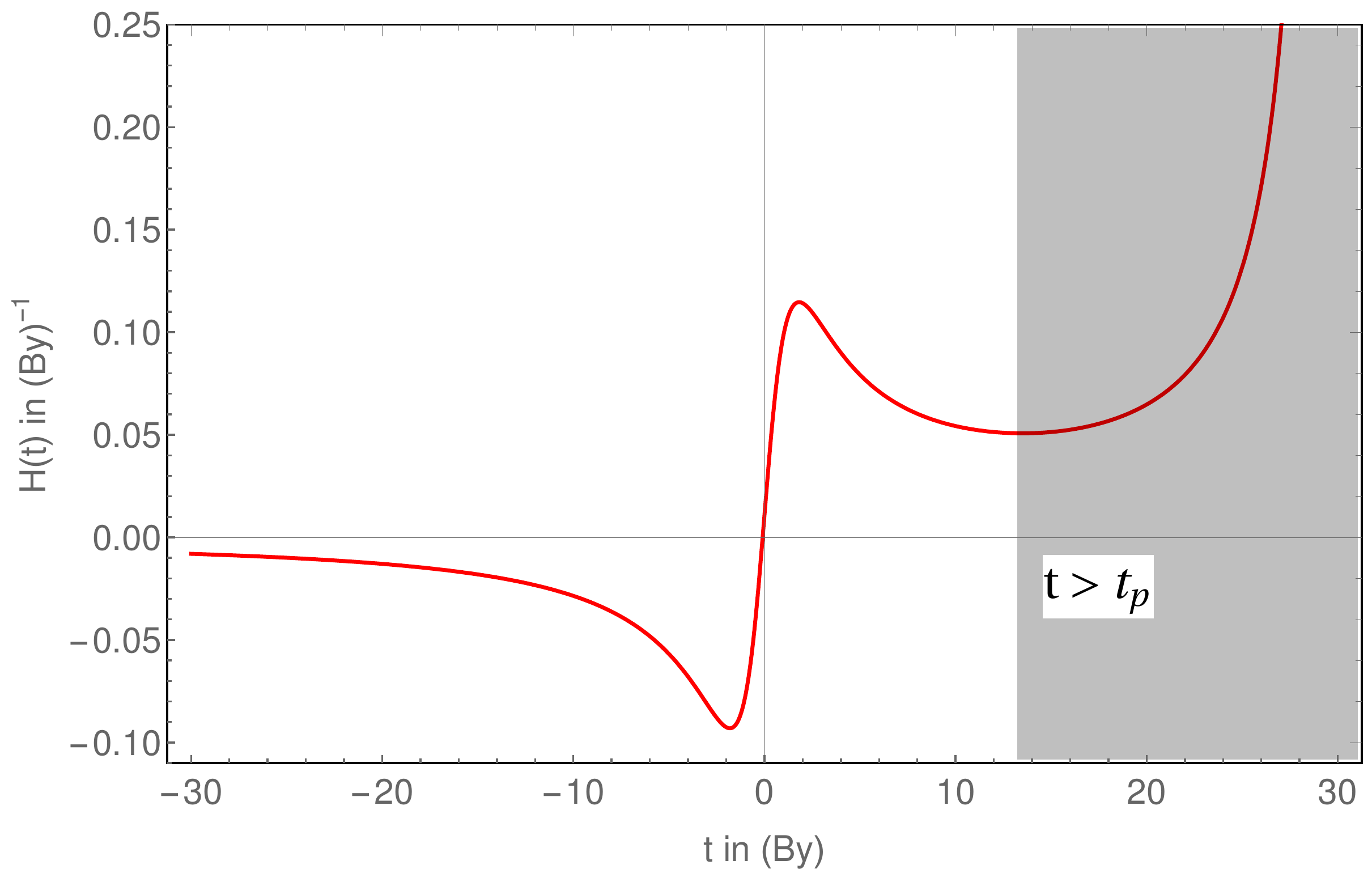}}~~
\subfloat[\label{Fig_2b}]{\includegraphics[scale=0.32]{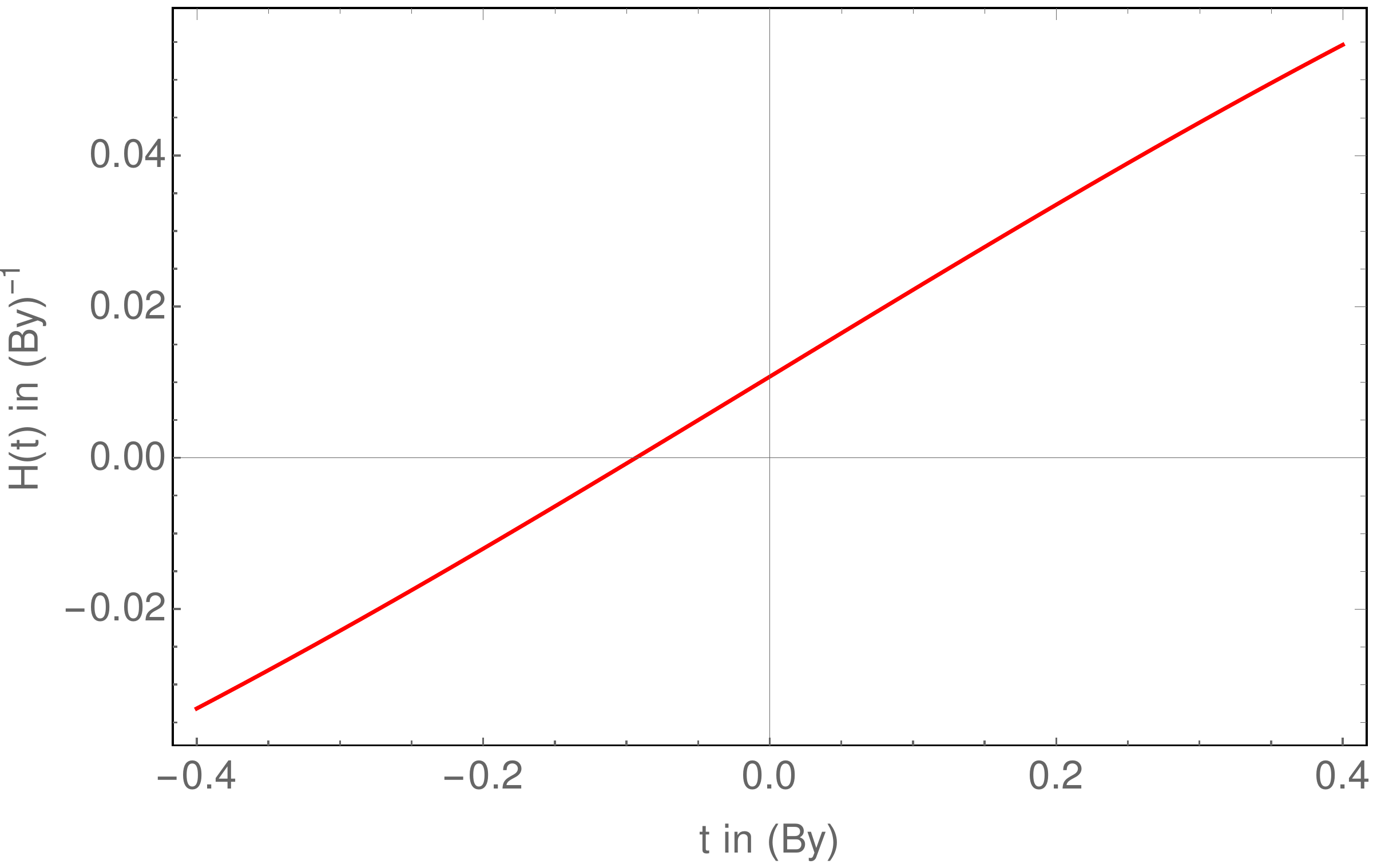}}
\caption{The above figure depicts the time evolution of (a) the Hubble parameter $H(t)$ and 
(b) the zoomed-in version of $H(t)$ near the bounce. Both the plots correspond to $n = 0.185$, $\alpha = 4/3$, $t_s = 30$ 
and $a_0 = 0.32$. Moreover the shaded region in the left plot corresponds to the cosmic time larger than the present age of the universe, 
i.e $t > t_p \approx 13.5\mathrm{By}$.}
\label{plot-Hubble}
\end{figure*}

\begin{figure*}[t!]
\centering
\subfloat[\label{Fig_3a}]{\includegraphics[scale=0.38]{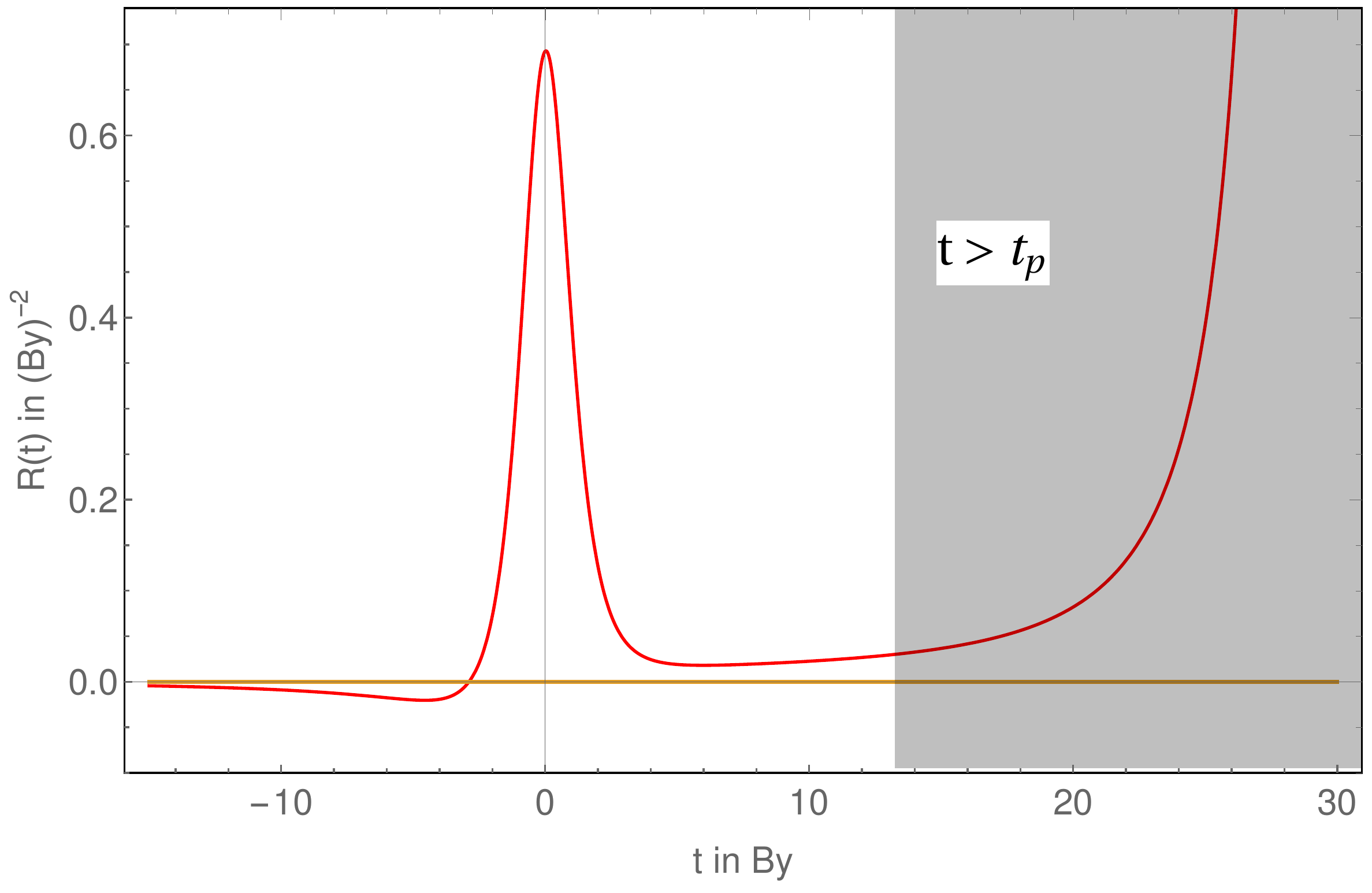}}~~
\subfloat[\label{Fig_3b}]{\includegraphics[scale=0.35]{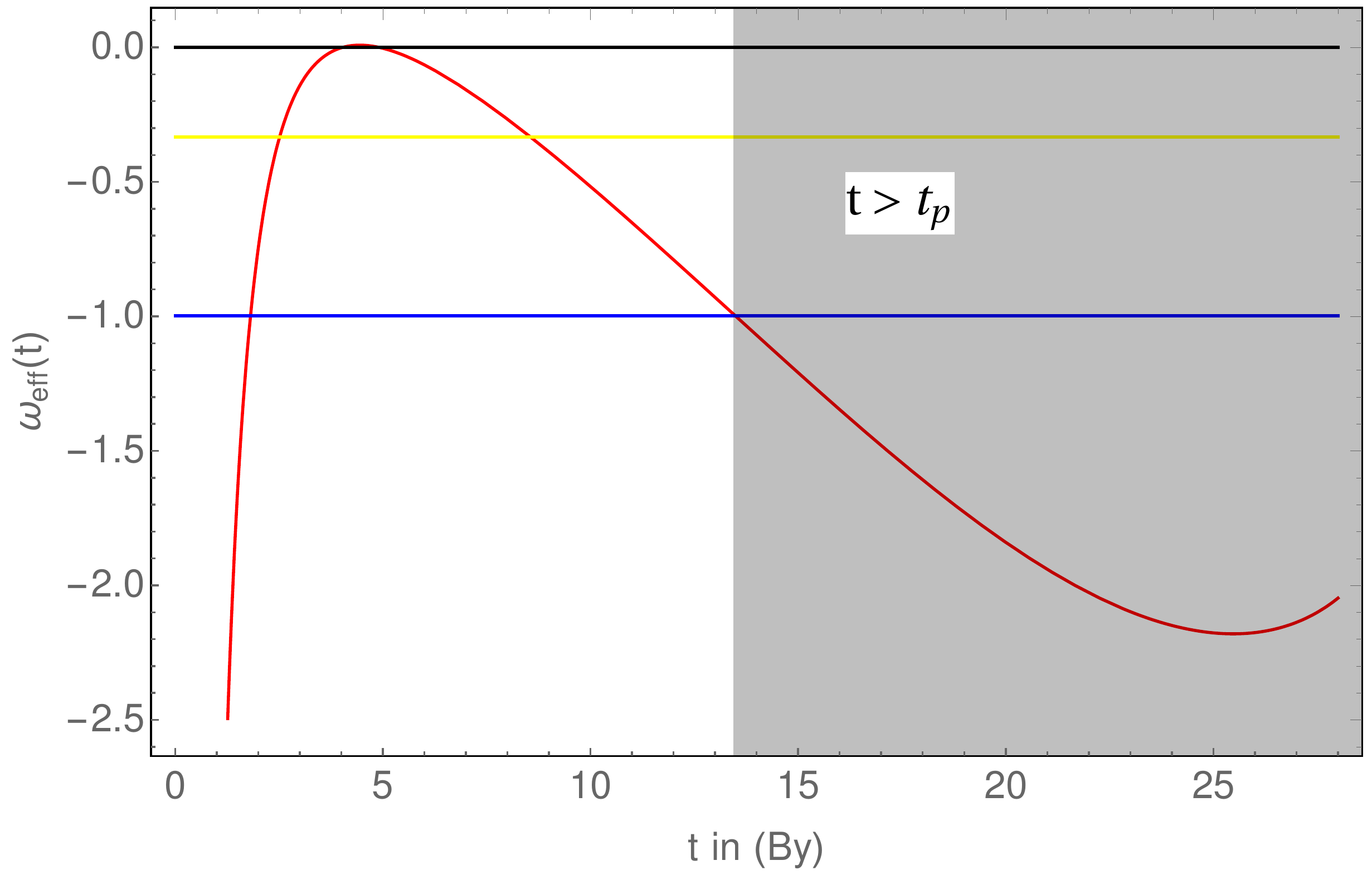}}
\caption{The above figure depicts the time evolution of (a) the Ricci scalar $R(t)$ and 
(b) the EoS parameter $\omega_\mathrm{eff}(t)$. Both the plots correspond to $n = 0.185$, $\alpha = 4/3$, $t_s = 30$ 
and $a_0 = 0.32$. Moreover the shaded region in the plots correspond to the cosmic time larger than the present age of the universe, 
i.e $t > t_p \approx 13.5\mathrm{By}$. In the right plot, the black, yellow and blue curve correspond to $\omega_\mathrm{eff} = 0,-\frac{1}{3},-0.997$ 
respectively. The curve $\omega_\mathrm{eff} = 0$ helps to investigate whether the present model exhibits a matter-like dominated epoch during some regime 
of cosmic time, the curve $\omega_\mathrm{eff} = -1/3$ is to demonstrate the accelerating or decelerating stages of the universe and the 
$\omega_\mathrm{eff} = -0.997$ curve reveals that the effective EoS of the present model matches with the Planck results at the present epoch i.e at 
$t = 13.5\mathrm{By}$.}
\label{plot-Ricci-EoS}
\end{figure*} 
 
 Keeping the parameter constraints in mind, we further give the plots of the background 
 $H(t)$, $R(t)$ and $\omega_\mathrm{eff}(t)$ (with respect to cosmic time) 
 by using Eq.(\ref{Hubble parameter}) and Eq.(\ref{ricci scalar}), see Fig.[\ref{plot-Hubble}] and Fig.[\ref{plot-Ricci-EoS}]. 
 The figures demonstrate -- (1) $H(t)$ becomes zero and shows an increasing behaviour with time near 
 $t \approx 0$, which indicates the instant of a non-singular bounce. (2) $R(t)$ starts from $0^{-}$ at asymptotic past. Moreover 
 the Ricci scalar gets a zero crossing from negative to positive values before the bounce occurs 
 and after that zero crossing the Ricci scalar seems to be positive throughout the 
 cosmic time. (3) The red curve of Fig.[\ref{Fig_3b}] represents the $w_\mathrm{eff}(t)$ for the present model while the yellow one of the same is for the constant 
 value $-\frac{1}{3}$. It is clear that $w_\mathrm{eff}$ exhibits two transitions from the early acceleration (near the bounce) to an intermediate deceleration 
 and then from the intermediate deceleration to the late time acceleration where 
 $\omega_\mathrm{eff}(t_p) \simeq -0.997$ consistent with the Planck-2018+SNe+BAO results \cite{Aghanim:2018eyx}. During the intermediate deceleration era, 
 $w_\mathrm{eff} \approx 0$ which indicates a matter dominated epoch. Such evolution of $w_\mathrm{eff}(t)$ 
 clearly reveals a smooth unification from a non-singular bounce to the dark energy era with an intermediate matter dominated stage.
  
\begin{figure*}[t!]
\centering
\subfloat[\label{Fig_5b}]{\includegraphics[scale=0.45]{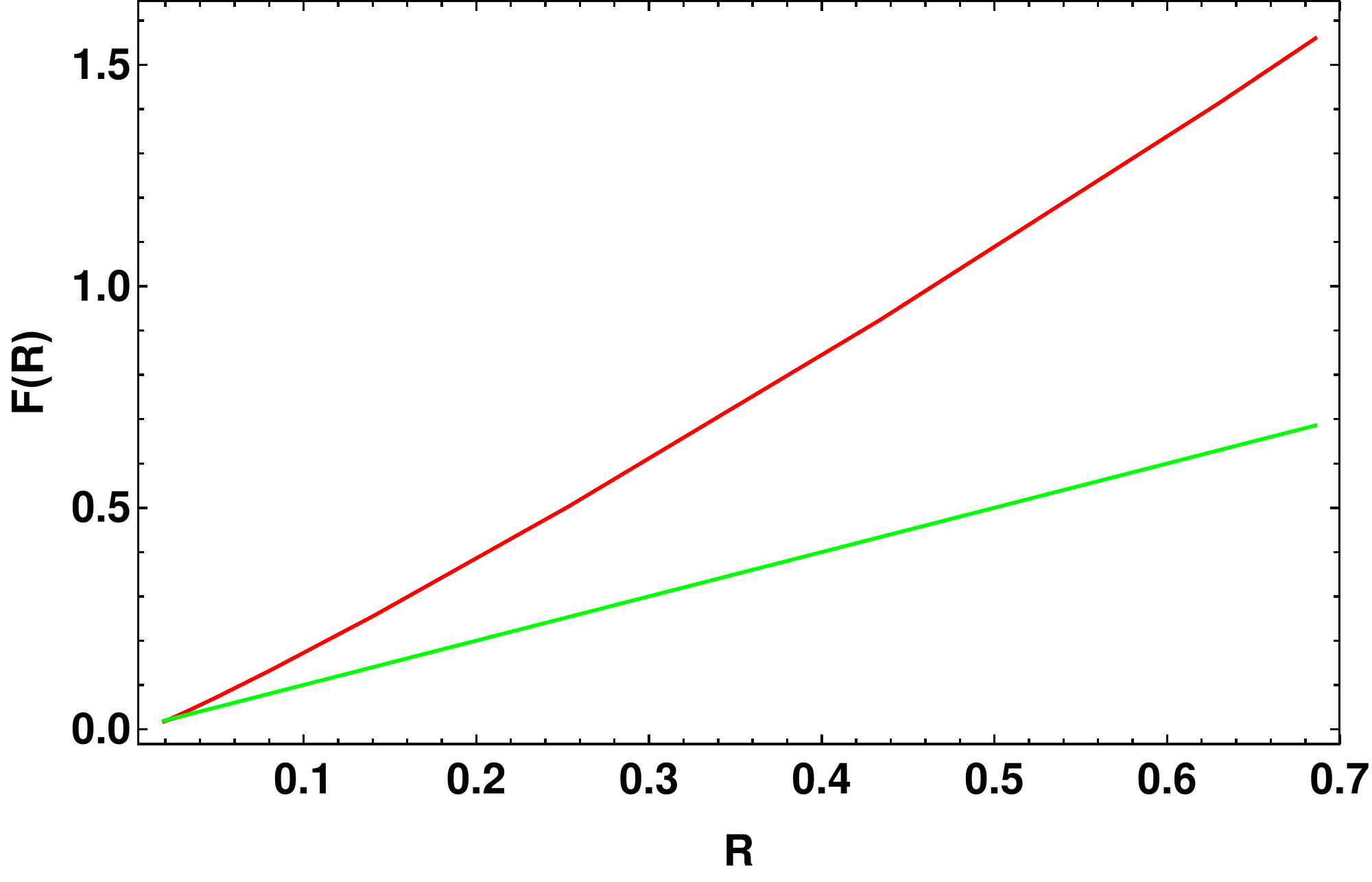}}
\caption{$F(R)$ vs. $R$. The initial condition during this numerical analysis is considered to be 
$F(R) = \left(R/R_0\right)^{\rho} + \left(R/R_0\right)^{\delta}$. 
The plot corresponds to $n = 0.185$, $\alpha = 4/3$, $t_s = 30$ and $a_0 = 0.32$.}
\label{plot-F-solution}
\end{figure*}

 The remaining task is to determine the form of $F(R)$ from the gravitational Eq.(\ref{basic4}). 
 In accordance the form of $H(t)$ in Eq.(\ref{Hubble parameter}), the F(R) gravitational equation may 
 not be solved analytically and thus we will solve it numerically. For this purpose, we use Eq.(\ref{basic4}). Moreover 
 the initial condition of this numerical analysis is considered to be $F(R) = \left(R/R_0\right)^{\rho} + \left(R/R_0\right)^{\delta}$, 
 i.e the analytic form of $F(R(t))$ during the late contracting era is taken as the initial condition of the numerical solution. 
 Consequently the numerically solved F(R) is shown in the Fig.[\ref{Fig_5b}[. 
 Actually the form of  $F(R)$ is demonstrated by the red curve, while the green one represents the Einstein gravity. 
Fig.[\ref{Fig_5b}] clearly depicts that the F(R) in the present context matches with the Einstein gravity as the Ricci scalar approaches to the present value, 
while the F(R) seems to deviate from the usual Einstein gravity, when the scalar curvature takes larger and larger values.\\

 \subsection{Cosmological perturbation}\label{sec-perturbation}

 In the present context, the comoving Hubble radius at distant past goes as $r_h \sim |t|^{1-2n}$. Therefore for 
 $n < 1/4$ (see the aforementioned condition $C1$), the Hubble radius diverges at 
 $t \rightarrow -\infty$ -- this makes the generation era of the primordial perturbation at the early contracting stage within the deep 
 sub-Hubble radius.\\

The scalar Mukhanov-Sasaki perturbation variable (symbolized by $v_k(\eta)$ in the Fourier space) follows the equation like,
\begin{align}
\frac{d^2v_k}{d\eta^2} + \left(k^2 - \frac{1}{z(\eta)}\frac{d^2z}{d\eta^2}\right)v_k(\eta)
= 0 \, ,
\label{sp5}
\end{align}
where the function $z(t)$ in the context of Chern-Simons F(R) gravity theory has the following form,
\begin{align}
z^2(t) = \frac{a^2(t)}{\kappa^2\left(H(t) + \frac{1}{2F'(R)}\frac{dF'(R)}{dt}\right)^2} \bigg\{\frac{3}{2F'(R)}\left(\frac{dF'(R)}{dt}\right)^2\bigg\}\, .
\label{sp3}
\end{align}
As mentioned earlier that the perturbation modes generate at deep contracting stage 
where the Hubble parameter and $F(R)$ follow Eq.(\ref{late contracting expressions}) and 
Eq.(\ref{FR solution-late contracting}) respectively. Thereby using such expressions of $H(t)$ and $F(R)$, we determine 
$z(t)$ as,
\begin{eqnarray}
z(t) = \left\{\frac{a_0^n|R_0|^n\left[12n(1-4n)\right]^n}{\kappa\left(R/R_0\right)^{n+1/2-\rho/2}}\right\}\times\left(\frac{P(R)}{Q(R)}\right)
\label{sp4}
\end{eqnarray}
where $P(R)$ and $Q(R)$ are defined as follows,
\begin{eqnarray}
P(R) = \left\{\frac{\sqrt{\rho}(1-\rho)
\left[1 + \frac{\delta(\delta-1)}{\rho(\rho-1)}
\left(\frac{R}{R_0}\right)^{\delta-\rho}\right]}{\left[1
+ \frac{\delta}{\rho}\left(\frac{R}{R_0}\right)^{\delta-\rho}\right]^{1/2}}\right\}~~~~~\mathrm{and}~~~~~
Q(R) = \left\{2n + \frac{(1-\rho)\left[1 + \frac{\delta(\delta-1)}{\rho(\rho-1)}
\left(\frac{R}{R_0}\right)^{\delta-\rho}\right]}
{\left[1 + \frac{\delta}{\rho}\left(\frac{R}{R_0}\right)^{\delta-\rho}\right]}\right\}~~.
\label{P}
\end{eqnarray}

To evaluate the function $z$ in terms of $\eta$, we need the functional form of $R = R(\eta)$ which is given by, 
\begin{eqnarray}
R(\eta) = \frac{1}{\eta^{2/(1-2n)}}\left\{\frac{12n(1-4n)}{|R_0|\left[a_0^n(1-2n)\right]^{2/(1-2n)}}\right\} \propto \frac{1}{\eta^{2/(1-2n)}}~~.
\label{sp6}
\end{eqnarray}
Consequently we get,
\begin{align}
z(\eta) \propto \left(\frac{P(\eta)}{Q(\eta)}\right)\times\eta^{\frac{2n+1-\rho}{1-2n}}
\label{sp67}
\end{align}
where $P(\eta) = P(R(\eta))$ and $Q(\eta) = Q(R(\eta))$, with $P(R)$, $Q(R)$ are given in Eq.(\ref{P}). The 
above expression of $z = z(\eta)$ yields the expression of $\frac{1}{z}\frac{d^2z}{d\eta^2}$, which is essential for the Mukhanov-Sasaki equation,
\begin{align}
\frac{1}{z}\frac{d^2z}{d\eta^2}=&\frac{\xi(\xi-1)}{\eta^2}
\left[1 + \frac{2\delta(\delta-\rho)\left\{(1-\rho)^2 + 2n(1+\rho-\delta)\right\}}
{\rho(1-\rho)(4n-\rho)(2n+1-\rho)}\left(\frac{R}{R_0}\right)^{\delta-\rho}\right]
\label{sp7}
\end{align}
with $\xi = \frac{(2n+1-\rho)}{(1-2n)}$. Due to $\delta - \rho > 0$, the 
term containing $\left(R/R_0\right)^{\delta-\rho}$ within the parenthesis of Eq.(\ref{sp7}) can be safely 
considered to be small during the late contracting era as $R \rightarrow 0$ at $t \rightarrow -\infty$. 
As a result, $\frac{1}{z}\frac{d^2z}{d\eta^2}$ becomes proportional to $1/\eta^2$ i.e., $\frac{1}{z}\frac{d^2z}{d\eta^2} = \sigma/\eta^2$ with,
\begin{align}
\sigma = \xi(\xi-1)\left[1 + \frac{2\delta(\delta-\rho)\left\{(1-\rho)^2 + 2n(1+\rho-\delta)\right\}}
{\rho(1-\rho)(4n-\rho)(2n+1-\rho)}\left(\frac{R}{R_0}\right)^{\delta-\rho}\right]~~.
\label{spnew}
\end{align}
which is approximately a constant in the era, when the primordial perturbation modes generate deep inside the Hubble radius. In effect, 
the Mukhanov-Sasaki Eq.(\ref{sp5}) can be solved as follows,
\begin{align}
v(k,\eta) = \frac{\sqrt{\pi|\eta|}}{2} \left[c_1(k)H_{\omega}^{(1)}(k|\eta|) +
c_2(k)H_{\omega}^{(2)}(k|\eta|)\right]\, ,
\label{sp8}
\end{align}
with $\omega = \sqrt{\sigma + \frac{1}{4}}$ and $c_1$ and $c_2$ are integration constants. 
The consideration of Bunch-Davies vacuum initially, leads to these integration constants as $c_1 = 0$ and $c_2 =1$ 
respectively. Therefore the power spectrum for the curvature perturbation in super-Hubble regime becomes,
\begin{align}
P_{\Psi}(k,\eta) = \left[\frac{1}{2\pi}\frac{1}{z|\eta|}
\frac{\Gamma(\omega)}{\Gamma(3/2)}\right]^2 \left(\frac{k|\eta|}{2}\right)^{3 - 2\omega}\, .
\label{sp10}
\end{align}

The tensor Mukhanov-Sasaki variable ($v_{\lambda}(k,\eta)$) has the following equation:
\begin{align}
\frac{d^2v_{\lambda}(k,\eta)}{d\eta^2}
+ \left(k^2 - \frac{1}{z_{\lambda}(\eta)}\frac{d^2z_{\lambda}}{d\eta^2}\right)v_{\lambda}(k,\eta) = 0~~,
\label{tp4}
\end{align}
where $\lambda = L,R$ represents the polarization index and $z_{\lambda}$ is given by,
\begin{eqnarray}
 z_L^2(t) = \left(\frac{1}{\kappa}\right)a^2(t)F'(R)\left\{1 - \frac{2\dot{\nu}(R)k}{aF'(R)}\right\}~~~~\mathrm{and}~~~~
 z_R^2(t) = \left(\frac{1}{\kappa}\right)a^2(t)F'(R)\left\{1 + \frac{2\dot{\nu}(R)k}{aF'(R)}\right\}~~,
 \label{ten per z}
\end{eqnarray}
with $\nu(R)$ being the CS coupling function. It s evident that the CS term 
has considerable effects on the tensor perturbation evolution, unlike to the case of vacuum F(R) model. 
Such difference of the tensor perturbation evolution 
between the CS corrected F(R) and the vacuum F(R) theory reflect on the primordial observable quantity, particularly on the tensor to scalar ratio.

We consider $\nu(R)$ (having the mass dimension [-2]) to be a power law form of the Ricci scalar, i.e
\begin{eqnarray}
 \nu(R) = \frac{1}{\left|R_0\right|(m+1)}\left(\frac{R}{R_0}\right)^{m+1}~~,
 \label{nu-form}
\end{eqnarray}
with $m$ being a parameter. By using this form of $\nu(R)$ and $R=R(\eta)$ from Eq.(\ref{sp6}), 
we evaluate $z_{\lambda}(\eta)$ and $\frac{1}{z_{\lambda}(\eta)}\frac{d^2z_{\lambda}}{d\eta^2}$ and these read,
\begin{align}
z_{L,R} \propto \left\{1 \mp \frac{4n}{\rho\left[12n(1-4n)\right]}\left(\frac{R}{R_0}\right)^{(\delta-\rho)(1-g)}\right\}\times\eta^{(2n+1-\rho)/(1-2n)}
\label{tpnew}
\end{align}
and
\begin{align}
\frac{1}{z_{L,R}}\frac{d^2z_{L,R}}{d\eta^2} = \frac{\xi(\xi-1)}{\eta^2}
\left\{1 \mp \frac{16n(\delta-\rho)(1-g)}{\rho(4n-\rho)\left[12n(1-4n)\right]}\left(\frac{R}{R_0}\right)^{(\delta-\rho)(1-g)}\right\}
\label{tp6}
\end{align}
respectively. Due to the fact that $\delta-\rho$ is positive, the variation of the term in the parenthesis in Eq.~(\ref{tp6}), 
can be regarded to be small in the low-curvature regime where the perturbation modes generate, and thus 
$\frac{1}{z_{\lambda}}\frac{d^2z_{\lambda}}{d\eta^2}$ becomes proportional to 
$1/\eta^2$ that is $\frac{1}{z_{\lambda}}\frac{d^2z_{\lambda}}{d\eta^2} = \sigma_{\lambda}/\eta^2$ (with $\lambda = L,R$), where
\begin{align}
\sigma_{L,R} = \xi(\xi-1)
\left\{1 \mp \frac{16n(\delta-\rho)(1-g)}{\rho(4n-\rho)\left[12n(1-4n)\right]}\left(\frac{R}{R_0}\right)^{(\delta-\rho)(1-g)}\right\}~~,
\label{tp9}
\end{align}
where we parametrize $m = \rho - 3 + (\delta - \rho)(1-g)$ in respect to a new parameter $g$, and 
recall $\xi = \frac{(2n+1-\rho)}{(1-2n)}$. The above expressions yield the tensor power spectrum, defined with the initial Bunch-Davies vacuum state, 
so we have,
\begin{eqnarray}
 P_{h}(k,\eta) = P_{L}(k,\eta) + P_{R}(k,\eta)
 \label{tp10}
\end{eqnarray}
with
\begin{eqnarray}
P_{L}(k,\eta) = \left[\frac{1}{2\pi}\frac{1}{z_L|\eta|}\frac{\Gamma(\Omega_L)}{\Gamma(3/2)}\right]^2 \left(\frac{k|\eta|}{2}\right)^{3 - 2\Omega_L}~~~~
\mathrm{and}~~~~
P_{R}(k,\eta) = \left[\frac{1}{2\pi}\frac{1}{z_R|\eta|}\frac{\Gamma(\Omega_R)}{\Gamma(3/2)}\right]^2 \left(\frac{k|\eta|}{2}\right)^{3 - 2\Omega_R}~~.
\label{tp11}
\end{eqnarray}
The factor $\Omega_{L,R} = \sqrt{\sigma_{L,R} + \frac{1}{4}}$ where $\sigma_{L,R}$ is defined in Eq.~(\ref{tp9}).

Now we can explicitly confront the model at hand with the latest Planck observational data \cite{Akrami:2018odb}, so we 
calculate the spectral index of the primordial curvature perturbations $n_s$ and the tensor-to-scalar ratio $r$, as follows,
\begin{align}
n_s =  4 - \sqrt{1 + 4\sigma}\, , \quad
r = \left. \frac{P_{h}(k,\eta)}{P_{\Psi}(k,\eta)}\right|_{h.c}~~,
\label{obs1}
\end{align}
where $P_{\Psi}(k,\eta)$ and $P_h(k,\eta)$ are obtained in Eq.(\ref{sp10}) and Eq.(\ref{tp10}) 
respectively, and the suffix 'h.c' denotes the horizon crossing 
instant when the mode $k$ satisfies $k = \left|aH\right|$. 

It may be noticed that $n_s$ depends on $n$, while $r$ depends on $n$ and $g$. The theoretical expectations of $n_s$ and $r$ 
get simultaneously compatible with the Planck 2018 data for $0.1845 \lesssim n \lesssim 0.1855$, with $g = 0.5$. 
On contrary, here we would like to mention that in the vacuum F(R) model, the observable quantities like $n_s$ and $r$ are not 
simultaneously compatible with the Planck results in the background of a non-singular bounce where $a(t) \sim t^{2n}$ during early contracting stage. 
In particular, the scalar and tensor perturbation amplitudes in the vacuum F(R) bounce model become comparable to each other and thus the tensor-to-scalar 
ratio comes as order of unity which is excluded from the Planck data. However, in the Chern-Simons corrected F(R) theory, the CS coupling function 
considerably affects the tensor perturbation evolution, keeping intact the scalar type perturbation with that of in the vacuum F(R) case. In effect, 
the tensor perturbation amplitude in the Chern-Simons generalized F(R) bounce model gets suppressed compared to the vacuum F(R) case, 
and as a result, the tensor-to-scalar ratio in the present context becomes less than unity and comes within the Planck constraints.

Thus as a whole, the Chern-Simons generalized F(R) gravity theory provides a smooth unification from a vaiable non-singular bounce to the dark energy era 
with an intermediate matter dominated like deceleration stage, and the DE EoS is found to be well consistent with the recent 
observations. Here we would like to mention that in regard to the background evolution, the effective EoS parameter at distant past is given by 
$\omega_\mathrm{eff} = -1+\frac{1}{3n}$ which is indeed less than unity due to the aforementioned range of $n$ that makes 
the observable quantities viable with the Planck results (in particular, $0.1845 \lesssim n \lesssim 0.1855$). In effect, the anisotropic energy density 
grows as $a^{-6}$ during the contracting era and thus the background evolution in the contracting stage becomes unstable to the growth of 
anisotropies, which is known as BKL instability. Thereby the present bounce model is not ekpyrotic in nature and thus suffers from the BKL instability. 
Therefore it is important to investigate whether an ekpyrotic bounce can be unified to the present dark energy era, and such unification has 
been proposed in Gauss-Bonnet theory of gravity by some of our authors in \cite{Nojiri:2022xdo}.

\section{\underline{Conclusion}}
Inadequecy of General Relativity and it's possible modification from different points of view  is a subject of interest for a long time.
Among these, most notable are gravity models in higher dimensions,  inclusion of higher curvature 
terms, asymmetric connection ( often referred to as space-time torsion ), Chern-Simons modified $F(R)$ gravity. 
All these features have their natural origin in the context of string theory.
In this work we have reviewed three different models in the light of a possible non-singular bounce each of which transcends Einstein's 
theory to incorporate new Physics and resolves some serious shortcomings of bounce cosmology. 
In the context of a generalized two brane warped geometry model , the resulting modulus field is known to acquire a potential from the brane vacuum energy which exhibits a metastable minimum. This potential results into a transient phantom epic which is shown to trigger a much desired bounce without the need of any other external field. In appropriate regime it yields a low scalar to tensor ratio  and thus making it consistent with Planck data. In the following discussion we have explored a higher curvature $F(R)$ gravity model in presence of a second rank anti-symmetric KR field. The third rank field strength of such a field is known to have geometric interpretation through the space-time torsion. Such  a model  which incorporates  higher curvature terms along with torsion is found to generate an ekpyrotic non-singular bounce. For appropriate choice of the parameter, the scale factor exhibits a non-singular bouncing scenario where  the spectral index for the curvature perturbation can be set to be consistent with Planck data. In yet another model which brings in the Chern-Somins coupled $F(R)$ gravity it is shown that such a coupling not only allows to have a non-singular symmetric bounce but also results into a smooth transition to dark energy era after an intermediate matter dominated decelerating phase of evolution. The hallmark of all the three different models is  each of them leads to a bouncing universe where the sources of the bouncing mechanism have  some underlying theoretical motivation and all 
of them provide a resolution of the singularity problem strictly within the domain of the classical cosmology.

\section{\underline{Data availability statement}}
Data sharing not applicable to this article as no data sets were generated or analysed during the current study.

\end{document}